\documentclass[twocolumn]{aastex63}

\hypersetup{linkcolor=red,citecolor=blue,filecolor=cyan,urlcolor=magenta}

\newcommand\halpha{H$\alpha$}
\newcommand\hbeta{H$\beta$}
\newcommand{\teff}{$T_{\rm eff}$}

\graphicspath{{./}{Figures/}}

\received{2021 June 16}
\revised{2021 July 22}
\accepted{2021 August 4}
\submitjournal{ApJ}

\shorttitle{Modeling Spica's \ion{H}{2} Region and Spectral Energy Distribution}

\shortauthors{}

\begin{document}

\title{Modeling the \halpha\ Emission Surrounding Spica
using the Lyman Continuum from a Gravity-darkened Central Star}

\correspondingauthor{Jason P. Aufdenberg}
\email{aufded93@erau.edu}

\author[0000-0002-4104-7580]{Jason P. Aufdenberg}
\affil{Embry-Riddle Aeronautical University \\
Physical Sciences Department \\
600 S. Clyde Morris Blvd\\
Daytona Beach, FL 32114, USA}

\author[0000-0002-6843-2673]{Joseph M. Hammill}
\affil{Embry-Riddle Aeronautical University \\
Physical Sciences Department \\
600 S. Clyde Morris Blvd\\
Daytona Beach, FL 32114, USA}



\begin{abstract}

The large, faint \halpha\ emission
surrounding the early B-star binary Spica has been used to constrain
the total hydrogen recombination rate of the nebula and indirectly
probe the Lyman continuum luminosity of the primary star. 
Early analysis suggested that a stellar atmosphere model, consistent
with Spica A's spectral type, has a Lyman continuum luminosity about
two times lower than required to account for the measured \halpha\
surface brightness within the nebula.
To more consistently model both the stellar and nebular emission,
we have used a model atmosphere for Spica A which includes the effects
of gravity darkening as input to  photoionization models to
produce synthetic \halpha\ surface brightness distributions for
comparison to data from the Southern \halpha\ Sky Survey Atlas (SHASSA).

This paper presents a method for the computation of projected surface
brightness profiles from 1D volume emissivity models
 and constrains both stellar and nebular parameters.
A mean effective temperature for Spica A of $\simeq$ 24,800 K
is sufficient to match both the observed absolute spectrophotometry,
from the far-UV to the near-IR, and radial \halpha\ surface brightness
distributions.  Model hydrogen densities 
increase with the distance from the star, more steeply and
linearly towards the southeast. 
The northwest matter-bounded portion of the  nebula is
predicted to leak $\sim17$\% of Lyman continuum photons. Model
\ion{H}{2} region column densities are consistent with archival
observations along the line of sight.

\end{abstract}

\keywords{Astronomy data modeling (1859), Diffuse nebulae (382), Ellipsoidal variable stars (455), Fundamental parameters of stars (555), Interstellar extinction (841), Spectral energy distribution (2129)}


\section{Introduction} \label{sec:intro}

The faint \halpha\ emission surrounding the two early B-type stars, classified as B1 III-IV and B2 V \citep{YBSC} and comprising the close binary (P = 4.01 days) \object{Spica} ($\alpha$ Virginis, HD 116658), 
provides a way to indirectly probe the Lyman continuum of early B-type stars, 
to test stellar atmosphere models 
and to potentially better constrain their supporting role in maintaining 
the warm-ionized medium in galaxies \citep{Haffner09}.  The Lyman continuum of hot stars, 
in the extreme ultraviolet (EUV), 
is generally hidden from direct observation, 
with rare exceptions, e.g. stars in the Canis Majoris tunnel
\citep{epsCMa,betCMa}, by bound-free 
absorption from neutral atomic hydrogen in the ground state along the line of sight.
Recent rocket-based observations
\citep{EpsCMaRocket2021}
are improving direct measurements of the EUV spectra of these 
B stars.

\begin{figure*}
    \centering
    \includegraphics[width=\textwidth]{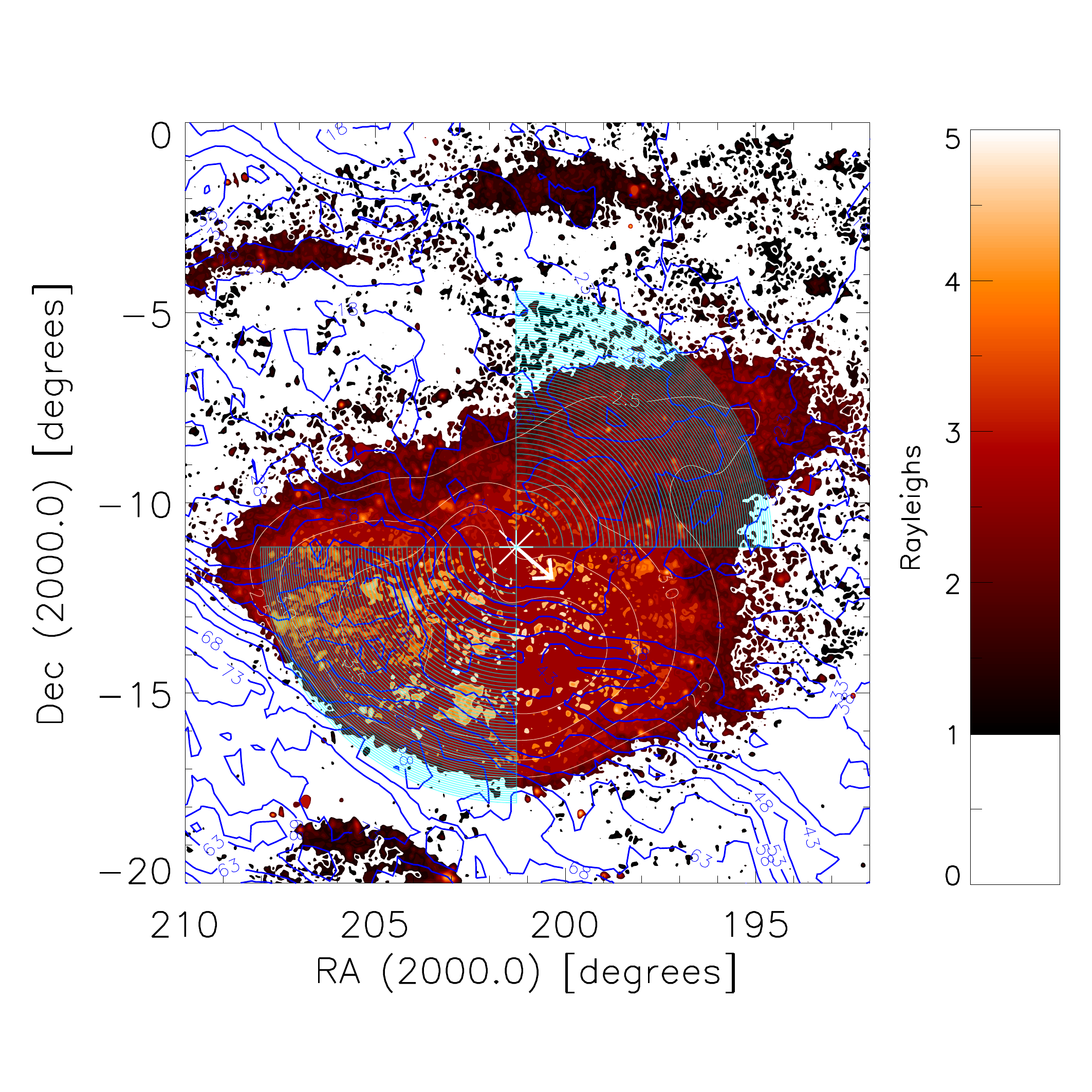}
    \caption{An \halpha\ surface brightness map around Spica using data from  SHASSA \citep{Gaustad_SHASSA,finkbeiner2003} after applying the stellar mask and transforming from galactic to equatorial
    coordinates. Filled contours show 15 levels from 1 R
     (Rayleighs $[\rm{R}]$, where $1\,\rm{R}=$ 2.4085$\times 10^{-7}$ erg cm$^{-2}$ s$^{-1}$ sr$^{-1}$ at \halpha)
    to 5 R and labeled orange contours from 2.5 R to 4 R are from the same map, but  Gaussian smoothed to 24 pixels (1$^\circ$). Velocity integrated \ion{H}{1} 21-cm column densities from the Leiden/Dwingeloo survey \citep{21cm} are the blue contours: 12 levels from 18 to 73 with a step size of 5 in units of $10^{19}\, {\rm cm^{-2}}$. The column densities are lowest in northwest corner and highest in the southeast corner.    Spica is marked near the center. The arrow denotes the direction  ($\mu_\alpha = -42.35\pm0.62\ {\rm mas\, yr^{-1}}, \mu_\delta =  -30.67\pm0.37\ {\rm mas\, yr^{-1}}$) and magnitude of the proper motion for 100,000 yr from Hipparcos \citep{V07}. Sixty quarter annuli to the southeast and northwest denote equal area regions were the median surface brightness was extracted for analysis, see Section \ref{sec:extraction}. }
    \label{fig:SHASSA_MAP}
\end{figure*}

The \halpha\ nebula is roughly $10^\circ$ in diameter on the sky (see Figure \ref{fig:SHASSA_MAP}). 
The first indirect  evidence
for ionized gas surrounding Spica came from a deficiency in neutral hydrogen
around the star based on 21-cm observations \citep{Fejes74}. \citet{york_kinahan} studied the ultraviolet interstellar absorption line spectrum towards Spica and estimated a radius
for the \ion{H}{2} region of 6.7 pc with electron density  $n_e \simeq 0.5\, {\rm cm^{-3}}$.
\citet[R85]{r85} detected the nebula in \halpha\ emission
and, based on these observations, estimated the total hydrogen recombination rate,  assuming a distance to Spica of 87 pc, and established a lower limit on the Lyman continuum luminosity,
$L_{\rm Ly} > 1.9\pm0.6 \times 10^{46}\ {\rm photons\, s^{-1}}$.  R85 and later \citep[R88]{r88}  found that the $L_{\rm Ly}$ prediction from a  B1 III stellar atmosphere model \citep{Panagia1973} with effective temperature, \teff\,=\,21,500 K,
in line with $22,400\pm1000\ {\rm K}$ from the interferometric angular diameter for Spica A
and photometry \citep{SSI1971}, was a factor of two lower than required, suggesting Spica was either significantly
closer (57 pc) or that the stellar parameters or models for Spica A were in need of revision.
The revised Hipparcos distance, $76.6\pm 4.1\, {\rm pc}$ from \citet{V07}, reduces the lower limit on
$L_{\rm Ly}$ by 20\%, but the discrepancy remains. 
There is a similar problem in a larger context: a high fraction (16/24) of the ``Lyman excess"  ultra-compact \ion{H}{2} regions identified by
\cite{Lyman_Excess_2018}, where the
ionization rate constrained by  the measurement of 5 Ghz free-free emission appears to exceed these stars' estimated Lyman continuum luminosities,
have central B0- and B1-type stars.

\cite{Park2010} used the much larger
\teff\,=\,26,000 K value for Spica A,  a value
consistent with the  \cite{Conti2008} temperature calibration for a 
B1 V star which is 3400 K hotter 
than the \cite{Panagia1973} calibration for the same spectral type,
to model radial \halpha\ surface brightness profiles of Spica's nebula. 
The nominal effective temperature
for a B1 III star is 24,000 K according to \citep{TSK1982},
only 21,500 K according \cite{Panagia1973}; 
\cite{Conti2008} does not provide a temperature calibration for
B giants. \cite{MOST_2016} found
a mean \teff = 25,300$\pm$500 K based on a detailed
spectroscopic analysis of Spica A, a value close to 24,700 K from a similar analysis by \cite{L1995},  while
\cite{Kunzli97} found \teff = 25791$\pm$464, based on the Geneva photometric system.
Effective temperatures for Spica A in the range 24,000 K to 26,000 K range are consistent with B1 III-IV based on recent effective temperature calibrations for non-supergiant B stars \citep{NP2014}, where B1.5 IV (EN Lac) is 23,000 $\pm$ 200 K, B1.5 III ($\alpha$ Pyx) is 22,900 $\pm$ 300 K and B1 IV ($\beta$ Cep) is 27,000 $\pm$ 450 K.

Furthermore, Spica A is a $\beta$ Cephei variable \citep{MOST_2016} and importantly an
ellipsoidal variable, both rotationally and tidally distorted, with a
hotter pole and cooler equator than the mean effective temperature,
making the apparent $L_{\rm Ly}$ a function of viewing angle.  

 The decrease in the effective temperature from
the pole to the equator is known as gravity darkening,
also known as von Zeipel darkening \citep{vz1,vz2},
expected for rapidly-rotating stars
to maintain both hydrostatic and radiative equilibrium.
Long-baseline interferometry of hot, rapidly-rotating stars
(see e.g., \cite{Che2011, vega2012}) provides direct confirmation 
of gravity darkening and  
indicates that the gravity darkening exponent, $\beta$, 
in the relation \teff $\, \propto g^\beta$, where $g$ is the local
surface gravity, is close to $\beta = 0.2$, compared to 
$\beta = 0.25$ for pure von Zeipel darkening (see Equation \ref{eqn:vonzeipel}). 
The theoretical work of \cite{ER2011} found that the $\beta$ parameter is a function of the flattening of the stellar pole: the more rapid the rotation, the lower the $\beta$ value, largely consistent with interferometric results.

In order to reconcile these estimates for the fundamental parameters of Spica A with the photoionization constraints from nebular observations, we have employed a stellar atmosphere model for Spica A which includes the effects of gravity darkening, as input to photoionization models  
for direct comparison to measured \halpha\ surface brightness distributions.

Section \ref{sec:spica_model} presents predictions from stellar atmosphere models
for the binary system. Section \ref{sec:extraction} describes the construction of mean \halpha\ surface brightness profiles to the southeast and northwest quadrants of the nebula from archival observations.
The construction of model surface brightness profiles from photoionization models is described in Section \ref{sec:comp_sb_section}.
Best fit models and parameter constraints are presented in Section \ref{sec:results}. Comparisons of our results to previous work are discussed in Section \ref{sec:Discuss}.

\section{Stellar atmosphere models for the Spica binary}\label{sec:spica_model}
We have used a model for the Spica binary system developed to 
fit interferometric, spectroscopic, and photometric data \citep{IAU2007}
which is described in detail in Appendix \ref{sec:appendixA}.
Unlike model atmospheres of gravity-darkened single stars (see e.g., \cite{Vega06}), the stellar components of close binary systems may be triaxial due to tidal distortion. Additionally the degree of
tidal distortion will be dependent on orbital phase for eccentric orbits. 

\begin{figure*}
    \centering
    \includegraphics[width=0.85\textwidth]{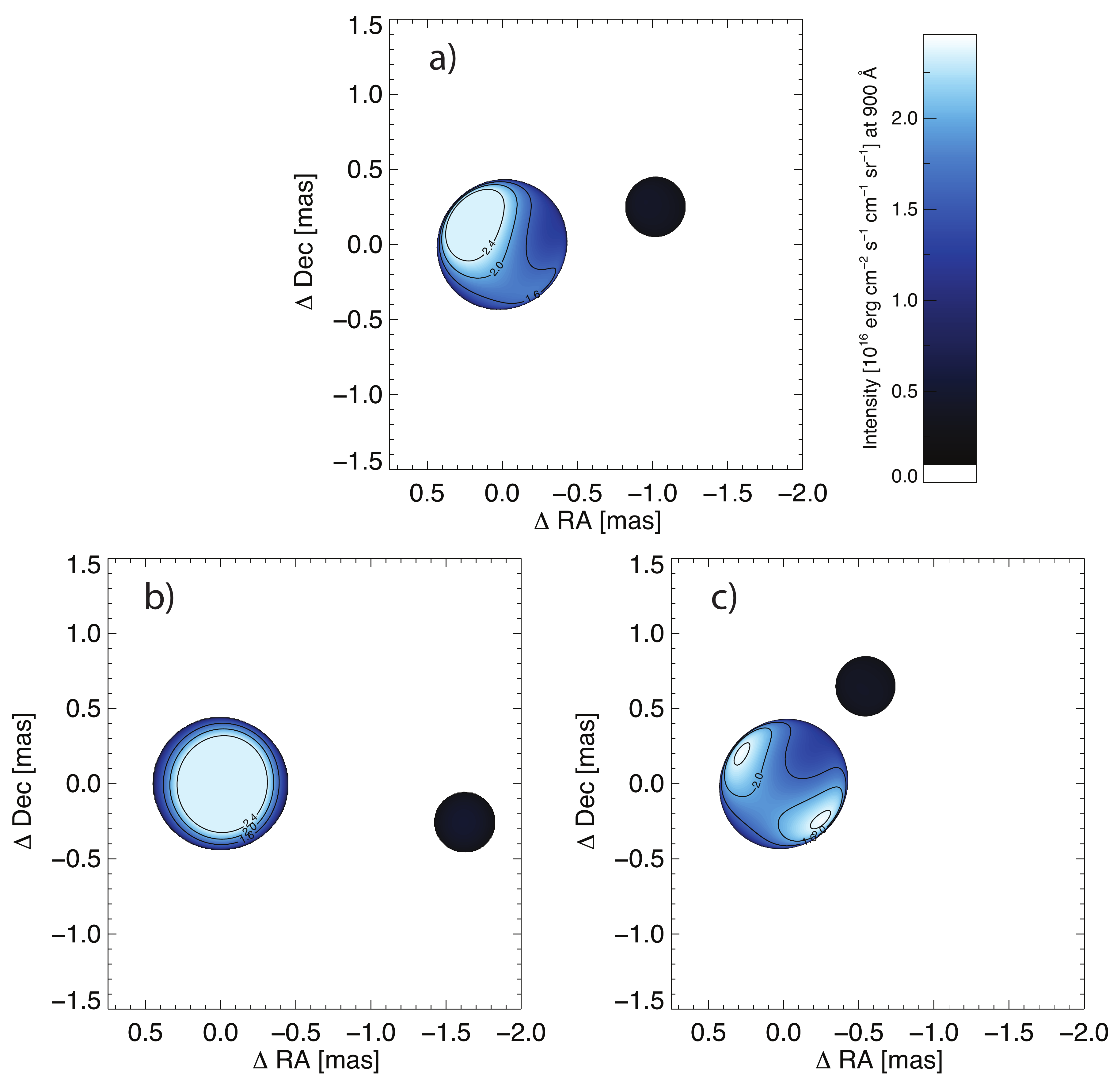}
    \caption{Monochromatic synthetic images of the Spica system in the Lyman continuum at 900\,\AA\ for three orbital inclinations: a) $i=116^\circ$ (view from Earth), b) 
    $i=180^\circ$ (pole-on) and  c) $i=90^\circ$ (edge-on), the stellar equatorial view.  The intensity contours are shown
    at (1.6, 2.0, 2.4) $\times 10^{16}\, {\rm erg\, cm^{-2}\, s^{-1}\, cm^{-1}\, sr^{-1}}$, however the maximum intensity does not reach the second contour for the $i=90^\circ$ view. Peak intensities for the three views are a) 2.46, b) 3.00 and c) 1.95 in the same units. The maximum intensity of the secondary star in the three views is 0.51 in b). Model parameters, based on a preliminary fit to interferometric, spectroscopic, and photometric data sets (Aufdenberg et al. 2021, in preparation):
    date, JD 2451544.5 (2000 January 1); the orbital period, $P = 4.0145898\, {\rm d}$, taken from 
    periastron to periastron; the ratio of the secondary mass to the primary mass, $q=0.6188$; the orbital eccentricity, $e = 0.123$;
    the epoch of periastron, $T_0= {\rm JD} 2440678.008$;
    the longitude of periastron at $T_0$, $\omega_0 = 136.727^\circ$;
    the apsidal period, $U= 111.886\, {\rm yr}$; 
    longitude of the ascending node, $\Omega = 309.938^\circ$;
   distance from Earth, $d = 78.514\, {\rm pc}$; semi-major axis, $a = 1.873\times 10^{12}\, {\rm cm}$;  
   surface gravity at the pole of the primary,  
   $\log_{10} (g_{\rm 1\ pole})= 3.741$; surface gravity at the pole of the secondary,  $\log_{10}(g_{\rm 2\ pole}) = 4.179$; 
   effective temperature at the pole of the primary,
   $T_{\rm eff\ 1\ pole} = 25642\, {\rm K}$ 
   (mean $T_{\rm eff\ 1}  = 24777\, {\rm K}$); 
   effective temperature at the pole of the secondary,
   $T_{\rm eff\ 2\ pole} = 22715\, {\rm K}$ 
(mean $T_{\rm eff\ 2}  = 22585\, {\rm K}$); 
non-synchronous rotation of the primary
$\omega_{\rm 1\ rot}/\omega_{\rm orb} = 1.8$;
non-synchronous rotation of the secondary
$\omega_{\rm 2\ rot}/\omega_{\rm orb} = 1.8$; 
gravity darkening exponent, $\beta = 0.205$. This model has
solar abundances \citep{A09}.}
    \label{fig:three_inclinations_of_spica}
\end{figure*}

Figure \ref{fig:three_inclinations_of_spica} shows synthetic images
of the binary in the Lyman continuum (at 900\, \AA) as viewed 
for different orbital inclinations, $i$: from Earth ($i=116^\circ$), pole-on ($i=180^\circ$),
 and edge-on ($i=90^\circ$), the stellar equatorial view.  
The peak in the intensity at the pole of Spica A is 54\% higher
in the pole-on view compared to the edge-on view.
Model parameter values for the orbital elements and stellar components
are based on preliminary simultaneous fits to interferometric, spectroscopic, and photometric data sets (Aufdenberg et al. 2021, in preparation), and are in generally good agreement the parameters from \cite{MOST_2016}. 

\begin{figure*}
    \centering
    \includegraphics[width=\textwidth]{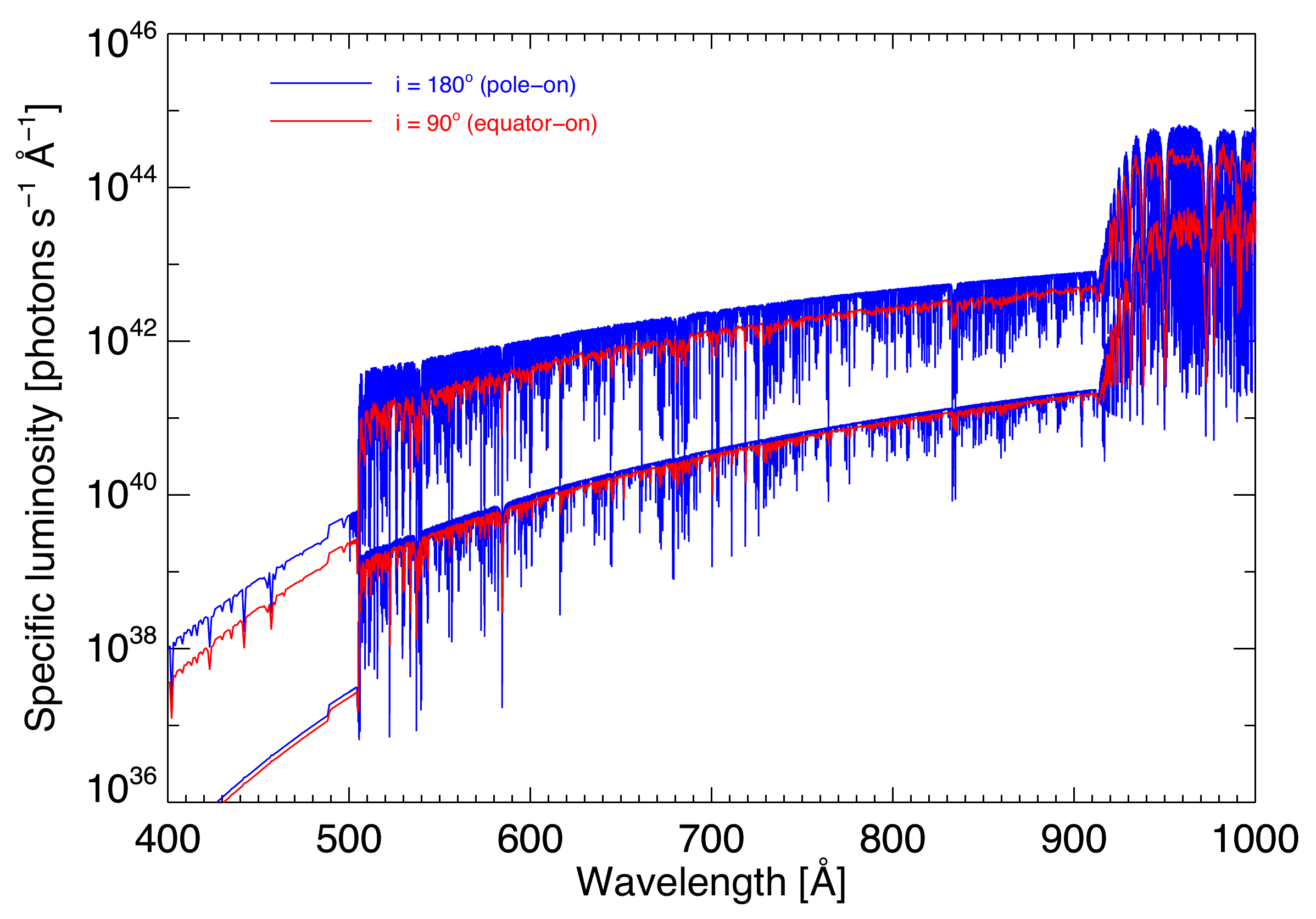}
    \caption{The predicted Lyman continuum for Spica A (above) and Spica B (below) for two inclinations: pole-on (blue) and equator-on (red). Model parameters given in the caption for Figure \ref{fig:three_inclinations_of_spica}. The equator-on spectra exhibit the greatest rotational line broadening, pole-on spectra exhibit no rotational broadening, hence the apparent difference in line blanketing for the two inclinations. The change in the appearance of the line blanketing at 500\,\AA\ is due to a change in the wavelength sampling from 1\,\AA\ below 500\,\AA\ to 0.05\,\AA\ above 500\,\AA.}
    \label{fig:lyman_continuum_sed}
\end{figure*}

Figure \ref{fig:lyman_continuum_sed} shows the specific luminosity,
%
\begin{equation}
    L_\lambda  =
    F_{1,2}(\lambda_{\rm obs}) \frac{\lambda d^2}{hc},
\end{equation}
(see Equation \ref{eqn:flux}) with units photons s$^{-1}$ {\AA}$^{-1}$,
in the Lyman continuum for both components
of the binary viewed pole-on and equator-on.  The primary star clearly
dominates the photon flux in the Lyman continuum. The more
compact Spica B shows less difference between the two inclinations than the larger, more distorted Spica A.

\begin{figure*}
\centering
\includegraphics[width=\textwidth]{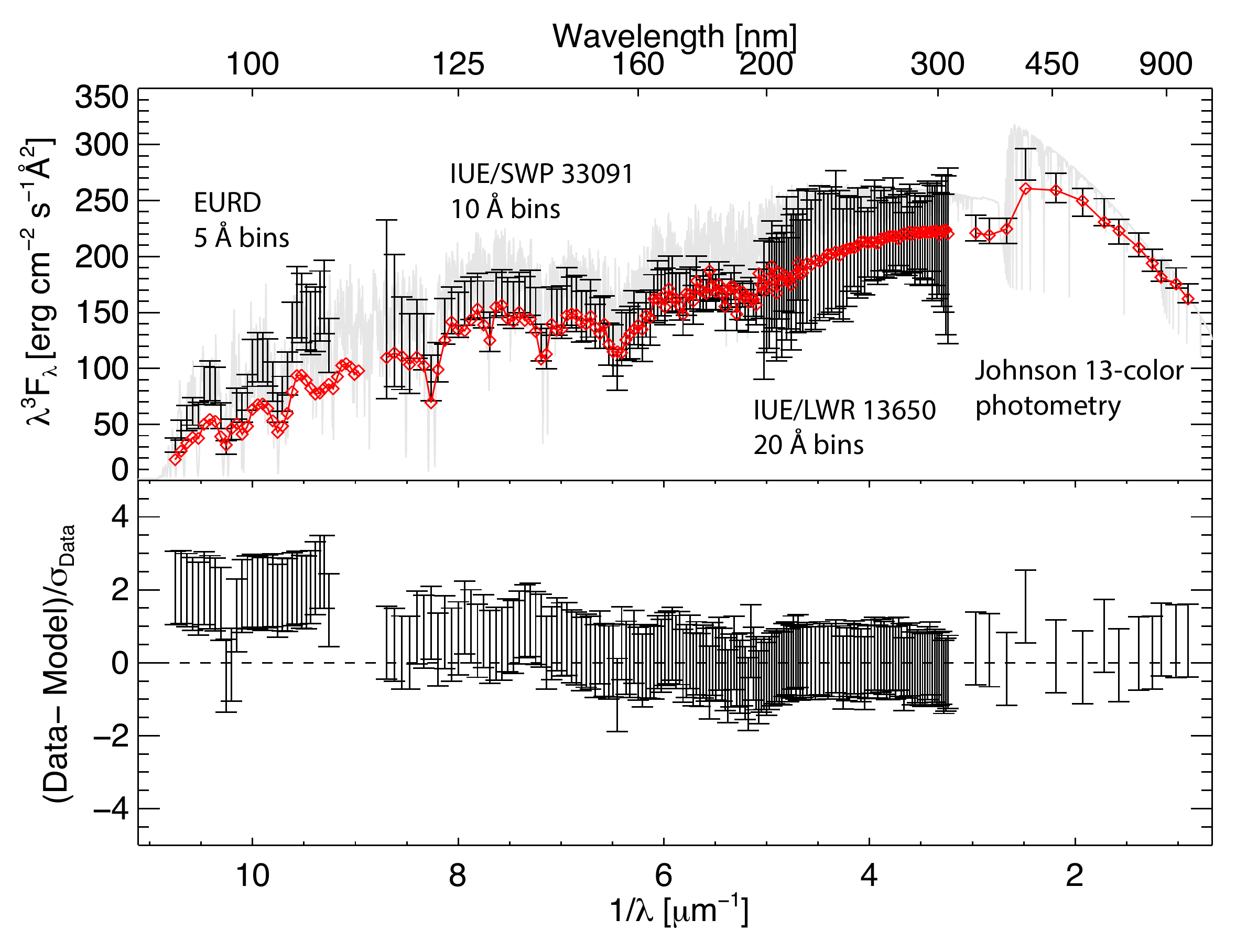}

    \caption{A model spectrum (red diamonds) after resolution reduction and
    binning (R $\simeq$ 40,000 model in light grey) compared 
    to observations from the Espectrógrafo Ultravioleta extremo para
    la Radiación Difusa     (EURD) spectrograph \citep{EURD} with a 
    resolution of R = 200 with 5\,\AA\ bins; International Ultraviolet Explorer (IUE)  with R=10,000, 10\,\AA\ bins (SWP camera) and R=15,000, 20\,\AA\ bins (LWR camera); 
    and 13-color spectrophotometry \citep{13color}.  Stellar model parameters given in the caption to Figure \ref{fig:three_inclinations_of_spica} for the $i = 116^\circ$ view.  This model uses the interstellar extinction curve from
    \cite{CCM} parameterized by the color excess, $E(B-V) = 0.0275$, and ratio of total of selective to visual extinction, $R(V) = 3.039$.} 
    \label{fig:spica_sed_comp}
\end{figure*}

The Lyman continuum luminosity from both stars is 
\begin{equation}
    L_{{\rm Ly}} = \int_0^{912\,\text{\AA}} 
    F_{\rm total} 
    \left(\frac{\lambda d^2}{hc}\right)\,d\lambda, 
\end{equation}
where $F_{\rm total}$ is defined in Equation \ref{eqn:flux_total}.
Table \ref{tab:lyman_fluxes} lists the predicted $L_{\rm Ly}$ values for a range of inclinations: 
the pole-on view Lyman luminosity is 1.6 times larger than the equator-on view.  The corresponding synthetic
spectral energy distributions are compiled in Table \ref{tab:SED_data}.  While
our current  best estimate for the gravity darkening exponent is $\beta \simeq 0.21$ (see Figure \ref{fig:three_inclinations_of_spica}),
we presently have no robust estimate for the uncertainly in this value.
Increasing $\beta$ to 0.25 (the classical von Zeipel value), while holding all other orbital and stellar parameters fixed, increases $\log_{10} L_{\rm Ly}$ 
most at $i=180^\circ$, by +0.03 (7\%), less at lower inclinations down to $i=90^{\circ}$, thus slightly increasing the contrast between between the pole-on and equator-on views.

\begin{deluxetable*}{CC}
\tablecaption{Predicted Lyman continuum luminosity from the surface of  Spica A as a function of the orbital inclination}
\label{tab:lyman_fluxes}
\tablehead{\colhead{Orbital inclination, $i$} &\colhead{$\log_{10} L_{\rm Ly}$}\\ \colhead{($^\circ$)} &\colhead{(photons s$^{-1}$ for $\lambda \le $ 912\,\AA)}}
\decimals
\startdata
90       &45.94 \\
100        &45.95 \\
110        &45.97 \\
116       &45.99 \\
120        &46.01 \\
130       &46.04 \\
140       &46.08 \\
150        &46.11 \\
160        &46.13 \\
170        &46.14 \\
180        &46.15 \\
\enddata
\end{deluxetable*}

\begin{deluxetable*}{RRRRRR}
\tablecaption{Synthetic spectral energy distributions for Spica A as a function of inclination}
\label{tab:SED_data}
\tablehead{\colhead{$\lambda$} 
&\colhead{$\log_{10} F_{\lambda{(i = 90^\circ)}}$}
&\colhead{$\log_{10} F_{\lambda{(i = 100^\circ)}}$}
&\colhead{$\log_{10} F_{\lambda{(i = 100^\circ)}}$}
&\colhead{$\log_{10} F_{\lambda{(i = 116^\circ)}}$}
&\colhead{$\log_{10} F_{\lambda{(i = 180^\circ)}}$}\\
\colhead{(\AA)} 
&\colhead{($\rm erg/cm^{2}/s/$\AA)} 
&\colhead{($\rm erg/cm^{2}/s/$\AA)} 
&\colhead{($\rm erg/cm^{2}/s/$\AA)} 
&\colhead{($\rm erg/cm^{2}/s/$\AA)} 
&\colhead{($\rm erg/cm^{2}/s/$\AA)}} 
\decimals
\startdata
 10.0 &-119.4098 &-119.2060 &-118.7955 &-118.4916  &-129.8959\\
 11.0 &-114.6337 &-114.4308 &-114.0008 &-113.6782  &-116.0968\\
 12.0 &-111.1009 &-110.8971 &-110.4489 &-110.1115  &-111.2714\\
\enddata
\tablecomments{Table \ref{tab:SED_data} is published in its entirety in the machine-readable format. A portion is shown here for guidance regarding its form and content.}
\end{deluxetable*}

The model spectral energy distribution (SED) of Spica for $i=116^\circ$ is
a pretty good match to archival absolute spectrophotometry
(\dataset[EURD and IUE data for Spica, HD 116658]
{http://svo2.cab.inta-csic.es/vocats/eurd/},
\dataset[13-color spectrophotometry data for Spica, HR 5056\ ]
{https://vizier.u-strasbg.fr/viz-bin/VizieR-5?-ref=VIZ60a459bc16acff&-out.add=.&-source=II/84/catalog&recno=742}) as shown in Figure \ref{fig:spica_sed_comp}, at least longward of 1100\, \AA. The
 model SED is systematically lower the Espectrógrafo Ultravioleta extremo para
la Radiación Difusa (EURD) spectrum \citep{EURD}, suggesting
perhaps a weaker far-UV extinction than provided by the \citep{CCM} mean extinction curve, a cooler primary star and/or an absolute calibration issue with the EURD data. We discuss these possibilities further in Section \ref{sec:Discuss}.

\section{Extracting Surface Brightness Profiles} \label{sec:extraction}

We chose to construct median \halpha\ surface 
brightness profiles in two directions: 
along the steepest gradient in 21-cm column density,\textbf{}
to southeast and to northwest of Spica (see Figure \ref{fig:SHASSA_MAP}). 
Surface brightness data from the Southern \halpha Sky Survey Atlas (SHASSA), \cite{Gaustad_SHASSA} via the compilation by \cite{finkbeiner2003}, were taken from files 
\verb|Halpha_map.fits| and \verb|Halpha_mask.fits|
(\dataset[SHASSA Data set]{https://faun.rc.fas.harvard.edu/dfink/skymaps/halpha/data/v1_1/index.html}). Those data with a bit-mask sum of 5  (SHASSA data + star
removal) were extracted from a $40^\circ \times 40^\circ$ region around Spica in galactic coordinates, followed by a transformation to equatorial coordinates.  Distances between any two
points in the map were computed on a sphere.  

The median \halpha\ surface brightness was computed in 60 quarter annuli bins of
approximately equal area to both the southeast and northwest (see Figure
\ref{fig:SHASSA_MAP}). Each bin contains between 450 and 500 pixels, except
for the innermost bin where 375 pixels remain after application of the star
removal mask.  In addition, very bright pixels were removed 
by 3$\sigma$ clipping about the median (using the IDL procedure
\verb|meanclip.pro|), at most 16 pixels in any bin.
We took the uncertainty in the surface brightness 
in each bin to be the standard deviation of brightness values plus any absolute 
difference between the mean and median values within the bin.
To the northwest, the mean and median of the surface brightness values differ by
less than 0.1 Rayleighs ($[\rm{R}]$, where $1\,\rm{R}=$ 2.4085$\times 10^{-7}$ erg cm$^{-2}$ s$^{-1}$ sr$^{-1}$ at \halpha) in all bins  except 10 
bins between 5.1$^\circ$ and 5.9$^\circ$ from Spica where the difference is as
large as 0.2 R. To the southeast, in all but 12 bins do the mean and
median surface brightness values differ by less than 0.1 R, however
between 6.6$^\circ$ and 6.9$^\circ$ from Spica the values differ by up to 0.3 R.

\begin{figure*}
    \centering
    \includegraphics[width=\textwidth]{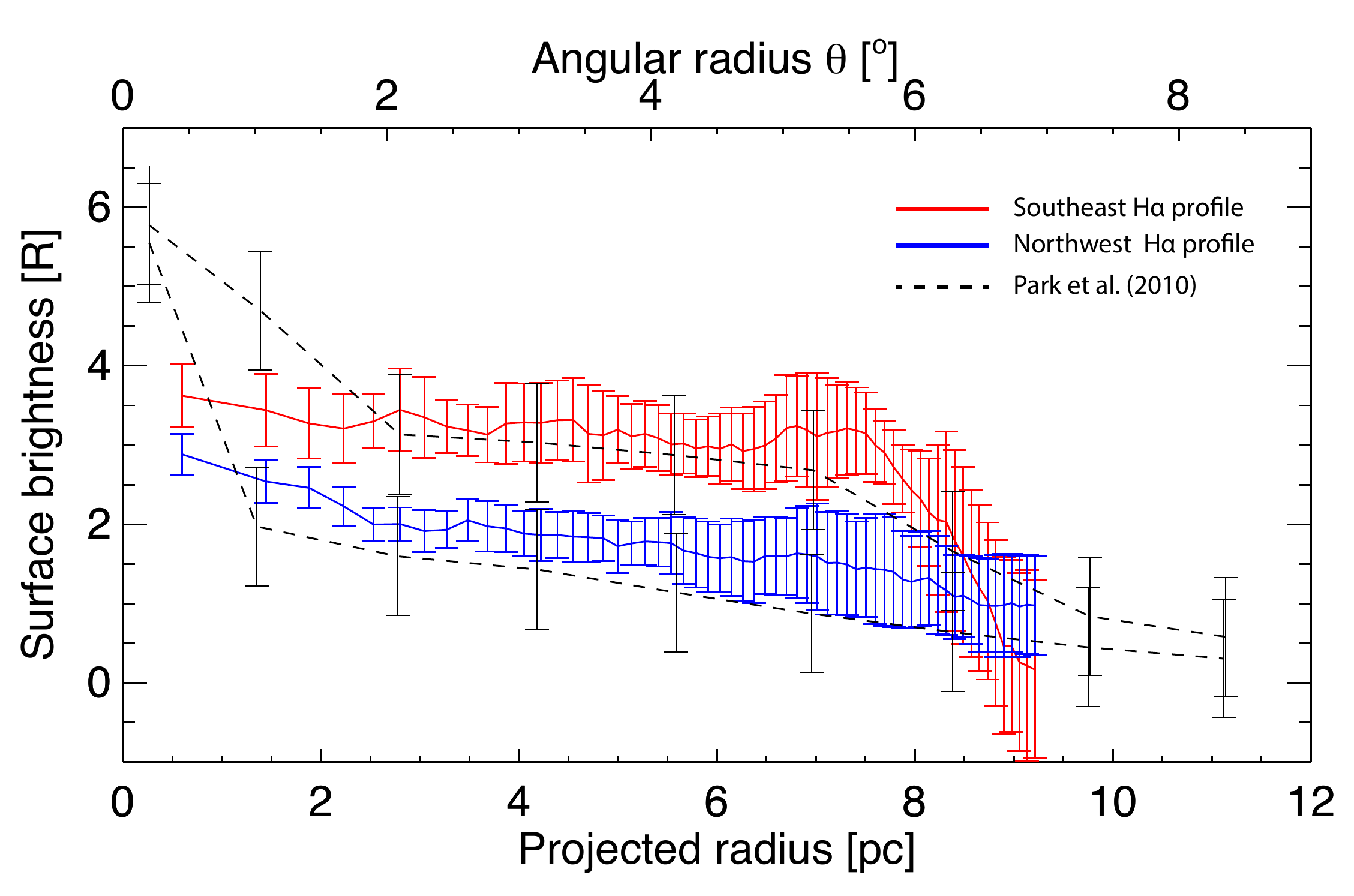}
    \caption{Median \halpha\ surface brightness profiles in 60 equal area quarter annulus bins (see Figure \ref{fig:SHASSA_MAP}) 
    towards southeast (red) and northwest (blue),
    compared to values extracted from \cite{Park2010} 
    for the whole south nebula (dashed, upper) and
    whole north nebula (dashed, lower).}  
    
   \label{fig:SB_profiles_park_comparison}
\end{figure*}

The histograms of the surface brightness in the outermost bins are clearly bi-modal indicative of
background component with a mode of $\sim 1.3$ R 
in the northwest and $\sim$ 0.7 R in the southeast.
This background emission likely originates from the warm-ionized medium of the Milky Way \citep{Haffner2010}.
We adopted an average background \halpha\ surface
brightness of 1.0 R which we subtracted from the median brightness.
The background-subtracted median surface brightness and uncertainty values for each bin are given in Table \ref{tab:mean_SB_values} and shown in Figure 
   \ref{fig:SB_profiles_park_comparison}.

\begin{deluxetable*}{ccDDDD}
\label{tab:mean_SB_values}
\tablecaption{Median \halpha\ surface brightness 
within quarter-annuli of equal area, extracted from the SHASSA dataset}
\tablehead{\colhead{Angular range \tablenotemark{a}} 
&\colhead{Projected radius \tablenotemark{b}} &\multicolumn{4}{c}{\halpha\ surface brightness \tablenotemark{c}}\\
& &\multicolumn{2}{c}{Northwest\tablenotemark{d}}
&\multicolumn{2}{c}{Southeast\tablenotemark{e}} \\
\colhead{($^\circ$)} &\colhead{(pc)} &\multicolumn{4}{c}{(Rayleighs)\tablenotemark{f}}}
\decimals
\startdata
  0.00 -- 0.89  &0.597     &2.882\pm  0.255 &3.621\pm  0.401 \\
  0.89 -- 1.26  &1.441     &2.540\pm  0.270 &3.439\pm  0.454 \\
  1.26 -- 1.55  &1.879     &2.460\pm  0.260 &3.271\pm  0.440 \\
  1.55 -- 1.79  &2.228     &2.228\pm  0.247 &3.207\pm  0.441 \\
\enddata
\tablecomments{Table \ref{tab:mean_SB_values} is published in its entirety in the machine-readable format.}

\tablenotetext{a}{Measured on a sphere between Spica and points within the nebula.}
\tablenotetext{b}{For the mean distance to Spica (76.6 pc) from \cite{V07}.}
\tablenotetext{c}{Sigma-clipped ($3\sigma$) median values and standard deviations with 
stellar mask applied and after background subtraction of 1.00 R.}
\tablenotetext{d}{Defined by equatorial coordinate constraints: 
$\delta > \delta_{\rm Spica}$ and $\alpha < \alpha_{\rm Spica}$.}
\tablenotetext{e}{Defined by equatorial coordinate constraints: 
$\delta < \delta_{\rm Spica}$ and 
$\alpha > \alpha_{\rm Spica}$.}
\tablenotetext{f}{At \halpha, 1 Rayleigh is 
$2.4085\times 10^{-7}\,{\rm erg\, cm^{-2}\, s^{-1}\, sr^{-1}}$.}
\end{deluxetable*}

\section{Computing model surface brightness profile} \label{sec:comp_sb_section}
\subsection{Models with no angular variation in the Lyman continuum luminosity} 
Given an input model SED\footnote{An example Cloudy input script and input stellar spectral energy distributions  are available in the Cloudy ``.ascii" format on-line}, the Cloudy code \citep[version 17.02]{Cloudy} returns a 1D radial profile of the \halpha\ volume emissivity 
($\varepsilon_{{\rm H}\alpha}$, units:
erg cm$^{-3}$ s$^{-1}$) as function of depth  into the nebula
from the inner radius $r_0$.
The nebula is assumed to be spherical, of radius $R_n$, and the volume emissivity has only a radial dependence. Figure \ref{fig:geometry}
shows the geometry used to integrate the volume emissivity along a set of rays, each corresponding to an angular radius $\theta$ measured from Spica in the plane of the sky, to compute a surface brightness distribution
($S$, units: Rayleighs $[\rm{R}]$).
The location $s$ along a given ray towards the observer is specified by the angle $\phi$ 
and the distance $r$ from the star.
$S$ is the integral of the
$\varepsilon_{{\rm H}\alpha}$ along each ray,

\begin{equation}
    S(\theta) = \frac{1}{4\pi}\int_{s(\phi_a)}^{s(\phi_b)} \varepsilon_{{\rm H}\alpha}(s)\, ds.
\label{eq:integral}
\end{equation}

The relationship between $s$, $\phi$ and
$r$ is

\begin{equation}
s(\phi) = 
    \sqrt{[r(\phi_a)\sin\phi_a -  r(\phi)\sin\phi)]^2 +
    [r(\phi_a)\cos\phi_a - r(\phi)\cos\phi]^2}.
\end{equation}
The starting and ending $\phi$ values are given by

\begin{equation}
\phi_a =  \pi - \theta - \sin^{-1}\left[\frac{d}{R_n}\sin\theta\right]\\
\end{equation}
\begin{equation}
\phi_b =  \sin^{-1}\left[\frac{d}{R_n}\sin\theta\right] - \theta\\
\end{equation}
 where $d$ is the distance of the star from Earth and $R_n$ is the outer radius of the nebula. 
 The endpoints can also be determined from the complex form of the $\sin^{-1}$, 
 \begin{equation}
     \phi_{a,b}=\mathrm{Re}\left[\frac{1}{i}\ln{\left(i\frac{d\sin{\theta}}{R_n}\pm\sqrt{1-\left(\frac{d\sin{\theta}}{R_n}\right)^2}\right)}\right].
 \end{equation}
 The radius $r$ for any
 $\phi$ and $\theta$ is 

\begin{equation}
r = \frac{d\sin\theta}{\sin(\theta+\phi)}.
\end{equation}

We chose 2500 and 400 discrete values for $\phi$ and $\theta$
respectively, using a cubic spline interpolation to find $\varepsilon_{{\rm H}\alpha}(r)$ at each corresponding $r$ value, and evaluating
Equation \ref{eq:integral} by numerical quadrature 
using a five-point Newton-Cotes integration formula.

 \subsection{Models with angular variation of Lyman continuum luminosity} \label{sec:inclinations}
\begin{figure*}
    \centering
    \includegraphics[width=\textwidth]{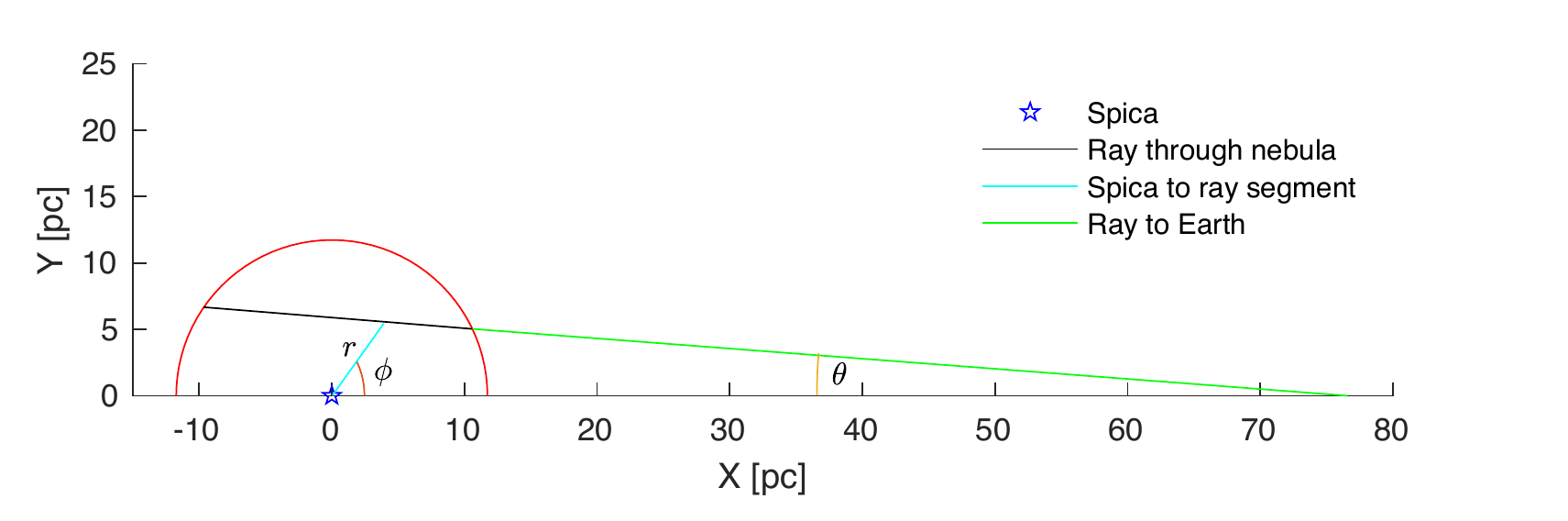}
    \caption{Depiction of the geometry used to define rays (lines of sight) through the nebula and predict a surface brightness distribution.  The angle $\theta$, with a vertex at Earth (located at X = 76.6 pc), measures the projected angular distance from Spica (at  X = 0), in the plane of the sky, for a ray through the nebula.  The angle $\phi$, with a vertex at Spica, is measured from the x-axis (directly towards the Earth) to points on a given ray of angle $\theta$ cutting through the nebula. The values of $\phi$ at the limits of each ray within the nebula are determined by the double-valued $\mathrm{arcsin}$ function. We integrated the volume emissivity (which is a function of the radius $r$ from a Cloudy model) over $\phi$ for each value of $\theta$ to find the surface brightness for each ray.}
    \label{fig:geometry}
\end{figure*}

  Building upon our model to integrate a single Cloudy model along rays through the nebula, we also developed a method
  to account for Spica A's temperature gradient in stellar colatitude, $\vartheta$ (see e.g., Equation \ref{eqn:lambda_cosine_radius_vector}), in a single surface brightness model.
     Each direction into the nebula (see Figure \ref{fig:geometry}) now has a different volume
     emissivity profile which is a function of $\phi$, corresponding to a value for $\vartheta$,
     the direction which in turn depends on the orientation of the binary described by the
     orbital inclination, $i$.  The corresponding $\vartheta$ at $\phi=0$ for a given $i$ is
     $\vartheta = 180^{\circ} - i$.
  A different atmosphere model was compiled for each $i$ in Table \ref{tab:lyman_fluxes}, and a
  Cloudy model computed for each.
  For the evaluation of Equation \ref{eq:integral}, we calculate values of $\varepsilon_{{\rm H}\alpha}(r,\phi)$ by a cubic spline interpolation of the Cloudy volume emissivity profiles first in $r$, then followed by $i$.

Any variation in the Lyman continuum  due to azimuthal distortions (e.g. along the line of apsides) of Spica A will be averaged out due to the short orbital
period relative to the characteristic hydrogen recombination timescale \citep[see their page 22]{agn3},
 \begin{equation}
\tau_{\rm rec} =\frac{1}{n_e\alpha_A} \approx 10^5/n_e\, {\rm yr} \simeq  3\times10^5\, {\rm yr}
 \end{equation}
taking the election density to be $n_e \simeq 0.3\, {\rm cm^{-3}}$ (see Section \ref{sec:results}), 
where  $\alpha_A$ is the  total recombination coefficient over all levels.
In addition, we find model $L_{\rm Ly}$ values to be largely
insensitive to orbital phase:  predictions for $L_{\rm Ly}$ at periastron (maximum tidal distortion)
are only 1\% lower relative to apoastron.
 This model therefore assumes Spica A's
temperature gradient is symmetric about its equator.
For example, the luminosity of the SED at an inclination of
$30^\circ$ is the same as that at an inclination of $150^\circ$.
We can also then perform rotations of $180^\circ$
about axes perpendicular to the binary orbital plane 
and recover the same SED.
  
\section{Model fitting and nebular constraints}\label{sec:results}
The gravity darkening of Spica A results in an SED and 
$L_{\rm Ly}$, which is 
a function of the viewing angle, here parameterized by the orbital inclination tabulated in
10$^\circ$ steps in Table \ref{tab:lyman_fluxes}.
For each inclination, the SED and associated $L_{\rm Ly}$ value were used
as 
inputs into a large grid of Cloudy models varying the inner cloud
radius, $r_0$, and the inner total hydrogen number density
at $r_0$, $n_0$, with the radial total hydrogen number density profile
described by a fixed power-law index, $\alpha$,
\begin{equation}
n(r)= n_0\left(\frac{r}{r_0}\right)^\alpha.    
\label{eqn:density_law}
\end{equation}

Cloudy does not permit varying $\alpha$ or SED (when the radiation field is specified by a user-defined \teff-log(g) grid) within a grid, so we ran separate grids for each combination of $\alpha$ and $L_{\rm Ly}$. Varying $L_{\rm Ly}$ is not sufficient to describe the variation in the SED, as this would simply scale an SED of identical shape for all inclinations.
The base set of models for both the northwest and southeast quadrants
included grids with $\alpha$ values from 0.0 to 1.4 in 0.1 steps, with
$L_{\rm Ly}$ corresponding to inclinations from 90$^\circ$ to 180$^\circ$ in $10^\circ$ steps.
Each grid was parameterized by $\log_{10} r_0$ values ranging from
17.50 to 19.50 in 0.02 dex steps (in units of cm) and $n_0$ values ranging from
0.010 to 0.400 in 0.005 steps (in units of cm$^{-3}$) resulting in 1,196,850
models in total for the base set. We computed
additional grids specifically for the northwest and southeast quadrants. 
The northwest quadrant specific grids contain smaller inner radii, $\log_{10} r_0$ from 16.50 to 17.48 in 0.02  dex steps, for an additional 59,250 models with same 
$n_0$, $\alpha$, $i$ ranges as the base set. The southeast quadrant specific 
grids contain larger $\alpha$ values: 1.5, 1.6, and 1.7,  
for an additional 239,370 models with the same $r_0$, $n_0$, $i$ ranges as the base set and a grand total of 1,495,470 Cloudy models.

For each Cloudy model we computed the corresponding surface brightness profile for a spherical
nebula from Equation \ref{eq:integral}.
Next we computed median values from these synthetic profiles in the same 60 equal area bins for direct comparison
to the extracted profiles for the northwest and southeast quadrants from the SHASSA data set (see Table \ref{tab:mean_SB_values}).
We first constrained the best-fit models by selecting those models
with $\chi^2 \le 56 $, a reduced chi-squared $\chi_\nu^2 \le 1$ 
for sixty bins and four degrees of freedom, corresponding to the four parameters.  
Next, following \cite{Andrae2010}, we identified 
the subset of these models where the residuals,
\begin{equation}
R_i =\left[ \frac{M_i - D_i}{\sigma_{D_i}}\right],
\end{equation}
are  consistent with a  normal distribution,
using Shapiro-Wilk test, keeping
models with $p > 0.05$, using the R function \verb|shapiro.test|,
where $M_i$ is the  model surface brightness and
$D_i$ the observed surface brightness in bin $i$. 

\begin{figure*}
    \centering
    \includegraphics[width=\textwidth]{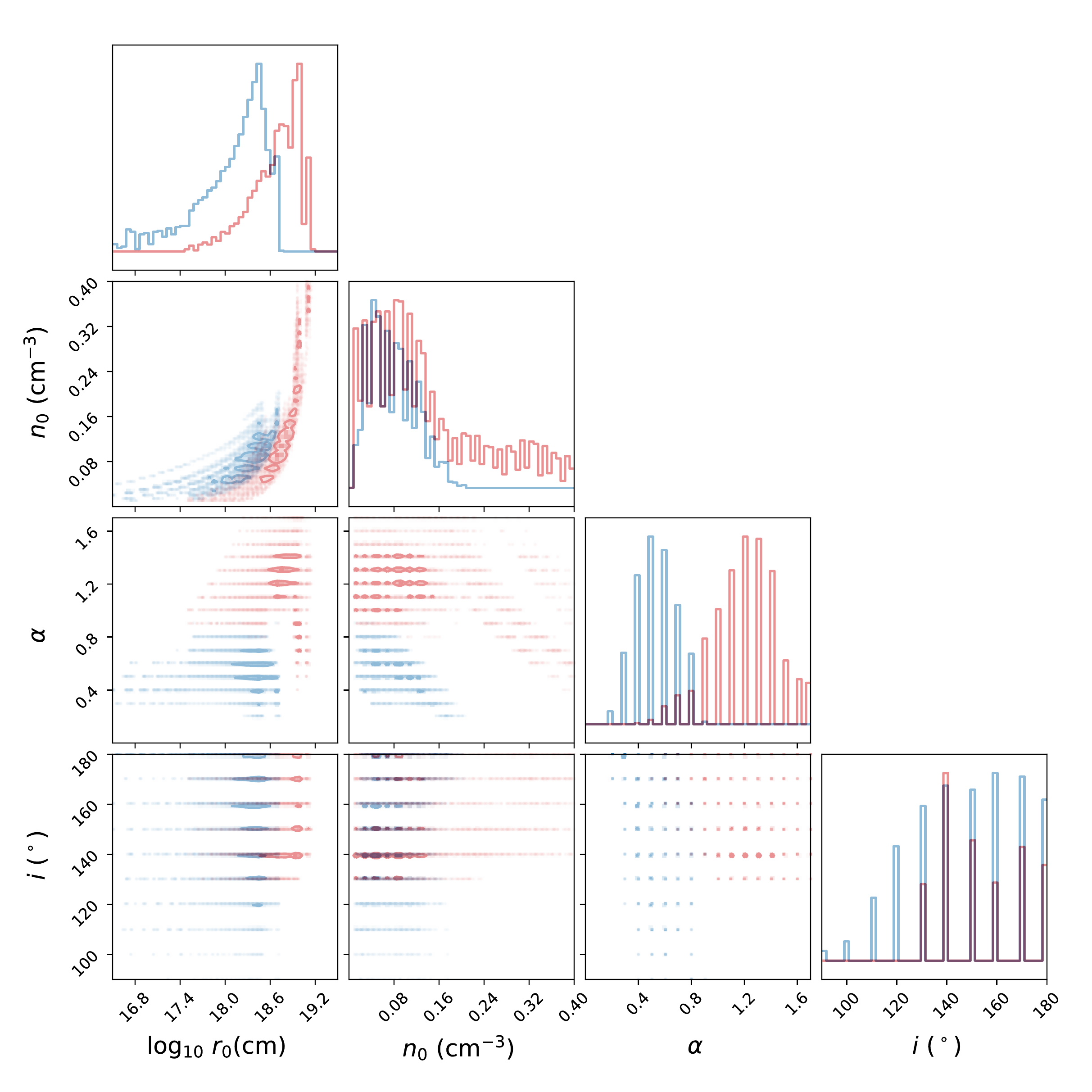}
    \caption{Distributions and correlations of best-fit parameters for the northwest (blue) and southeast (red) quadrants of the nebula.  The models
    selected to have a $\chi^2 \le 56$ with a distribution of residuals consistent with normality ($p > 0.05$) using the Shapiro-Wilk test. These critera
    yield 3615 models for the northwest and 3048 models for the southeast. The two quadrants differ most noticeably in the density power-law index $\alpha$ and in the threshold Lyman continuum value for the southeast, inclination $i \ge 130^\circ$.
    Plot generated using the Python package corner.py \citep{corner_plot}.}
    \label{fig:best_fit_north_south_independent}
\end{figure*}

Applying these criteria, 3615 models for the northwest quadrant 
and 3048 models for the southeast quadrant remain.  The 
distributions and correlations of the four parameters 
from these models are shown in Figure \ref{fig:best_fit_north_south_independent}.
While the parameters for northwest and southeast quadrants show
considerable overlap in $r_0$, $n_0$, and $i$, there is considerably
less overlap in the $\alpha$ values, justifying the separate analysis for
each quadrant. 

\begin{figure*}
    \centering
    \includegraphics[width=\textwidth]{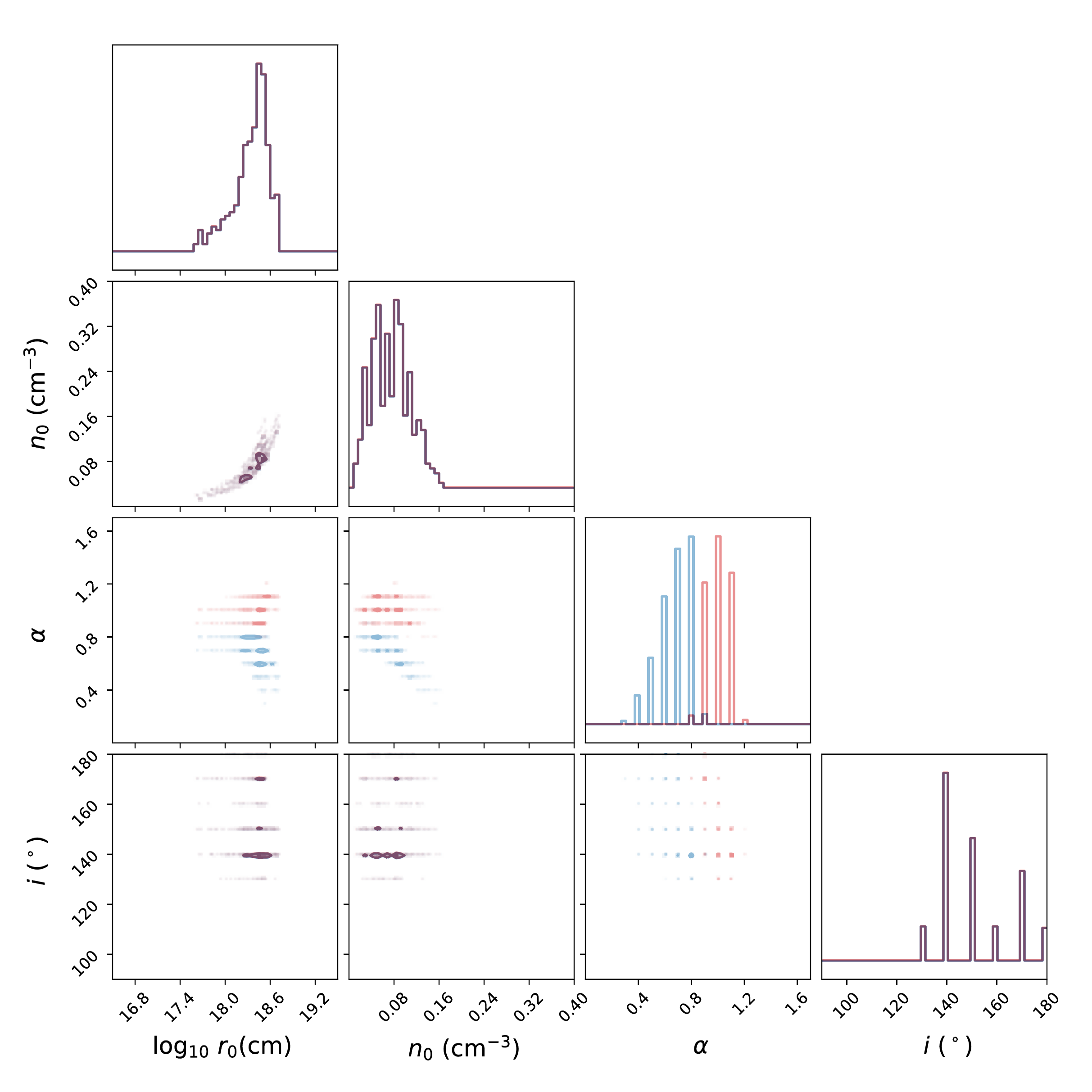}
    \caption{A subset of the models shown in Figure
    \ref{fig:best_fit_north_south_independent}, these 352 pairs of models for
    the northwest (blue) and southeast (red) quadrants have matching $r_0$,
    $n_0$, and $i$
    parameters  (hence the distributions of these parameters for both quadrants fall on top of each other), meaning the inner cloud radius, total
    hydrogen number density and Lyman continuum match on opposite sides of Spica.
    Plot generated using the Python package corner.py \citep{corner_plot}.}
    \label{fig:best_fit_north_south_match}
\end{figure*}

\begin{figure*}
    \centering
    \includegraphics[width=\textwidth]{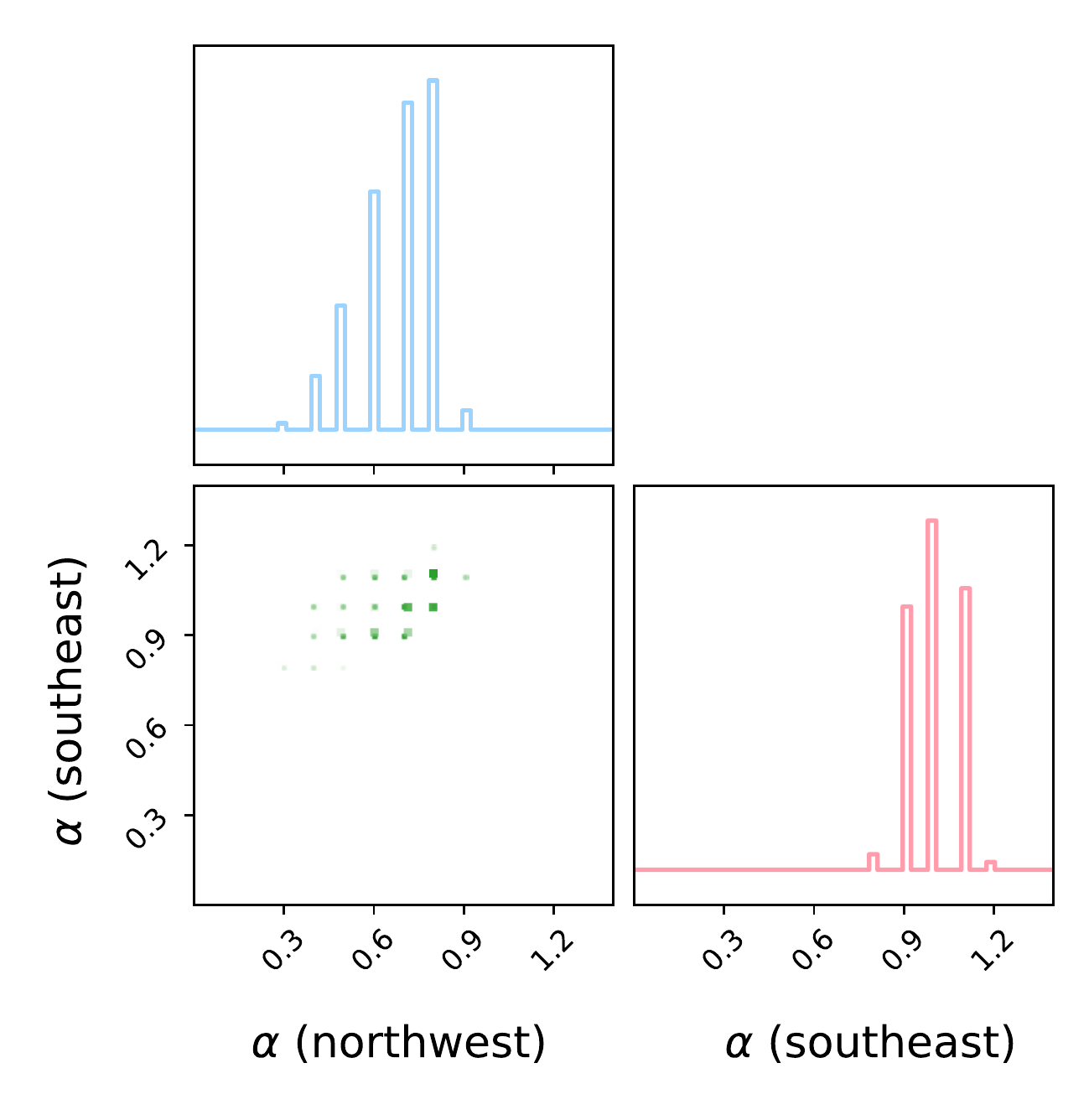}
    \caption{The distributions and correlations of density profile power-law index $\alpha$ for the 352 model pairs for the northwest and southeast quadrants (see Figure \ref{fig:best_fit_north_south_match}).  There is a weak positive correlation between the northwest and southeast $\alpha$ values when the models are forced to have the same inner radius, initial density, and Lyman continuum.}
    \label{fig:alpha_correlation}
\end{figure*}

Next we sought pairs of models from the northwest and
southeast quadrants with matching $r_0$ and $n_0$ values to force the
inner radius of the cloud and the hydrogen number density at the inner radius to match 
close to Spica. We further constrained the model pairs to have the same $i$
values such that two quadrants will see effectively equal and opposite views on
Spica A with the same Lyman continuum, $L_{\rm Ly}$. These additional constraints 
leave 352 pairs of models. Their parameters are shown in Figure
\ref{fig:best_fit_north_south_match}.  The range of $\alpha$ 
values is reduced relative to the independent fits shown in 
Figure \ref{fig:best_fit_north_south_independent}.  The $\alpha$
values for each northwest/southeast model pair are weakly correlated
as show in Figure \ref{fig:alpha_correlation}.

\begin{figure*}
    \centering
    \includegraphics[width=\textwidth]{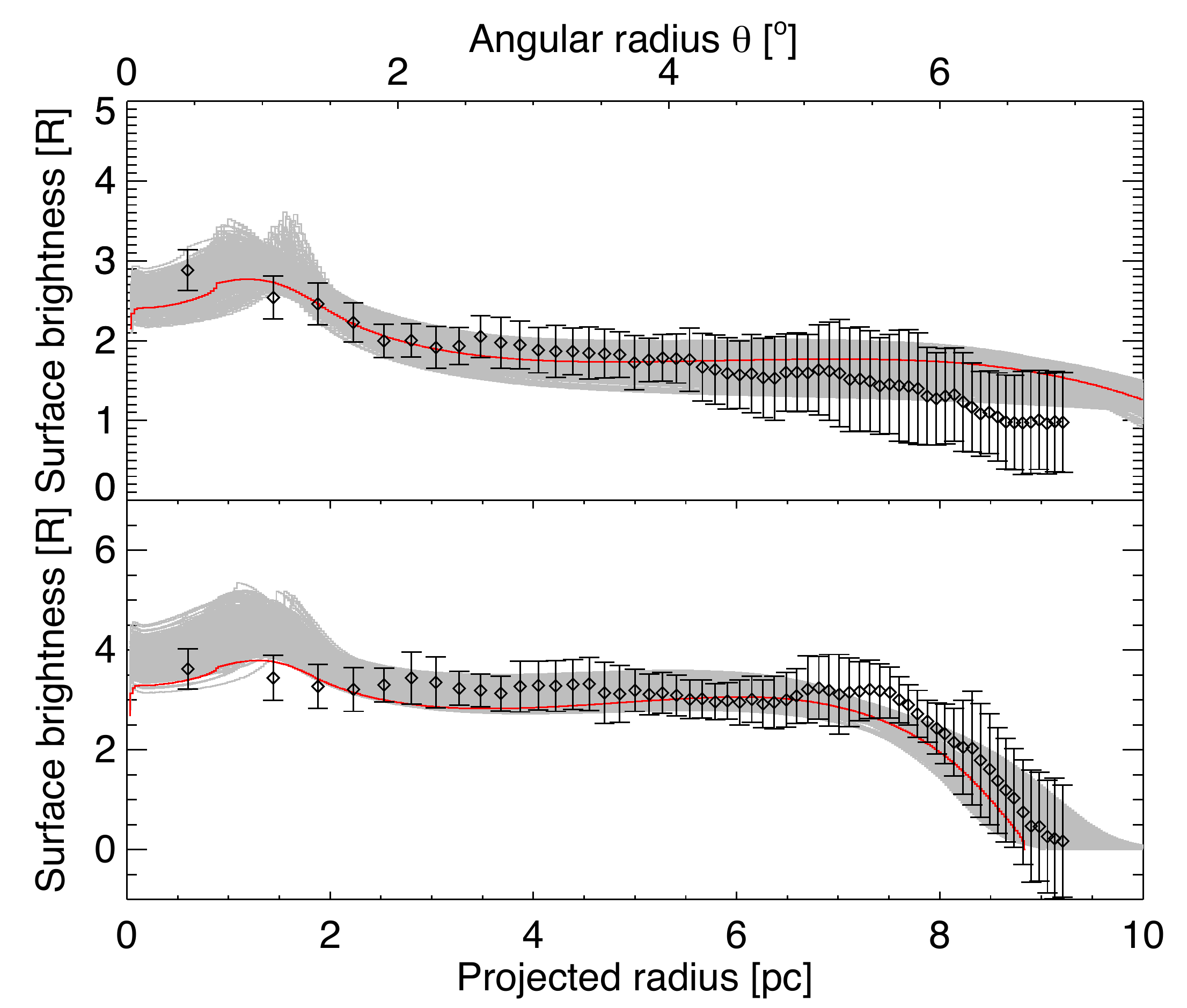}
    \caption{Synthetic \halpha\ surface brightness profiles (in grey), compared to binned surface
    brightness values from observations (see Table \ref{tab:mean_SB_values}),  from 352
    models for the northwest quadrant (above) and the southeast quadrant (below).
    The ranges for the parameters of these 352 model pairs are shown in Figure
    \ref{fig:best_fit_north_south_match}.  The model profiles in red correspond to a more physically
    realistic scenario where the Lyman continuum luminosity 
    varies as function of $\phi$ (see Figure \ref{fig:geometry} and Section
    \ref{sec:inclinations}) 
     such that different locations within the spherical model nebula see different projections of Spica
    and hence different Lyman continua (see Table \ref{tab:lyman_fluxes}), while from Earth we see the
    orientation $i = 116^\circ$ (see Figure \ref{fig:three_inclinations_of_spica}).
    This pair of models (in red) have $\log_{10} r_0\, [{\rm cm}] = 18.44$ 
    and $n_0 = 0.07\, {\rm cm^{-3}}$ with $\alpha = 0.8$ for the northwest (above)
    and $\alpha = 1.1$ for the southeast (below).}
     \label{fig:sb_profiles_for_matching_pairs}
\end{figure*}

\begin{figure*}
    \centering
    \includegraphics[width=\textwidth]{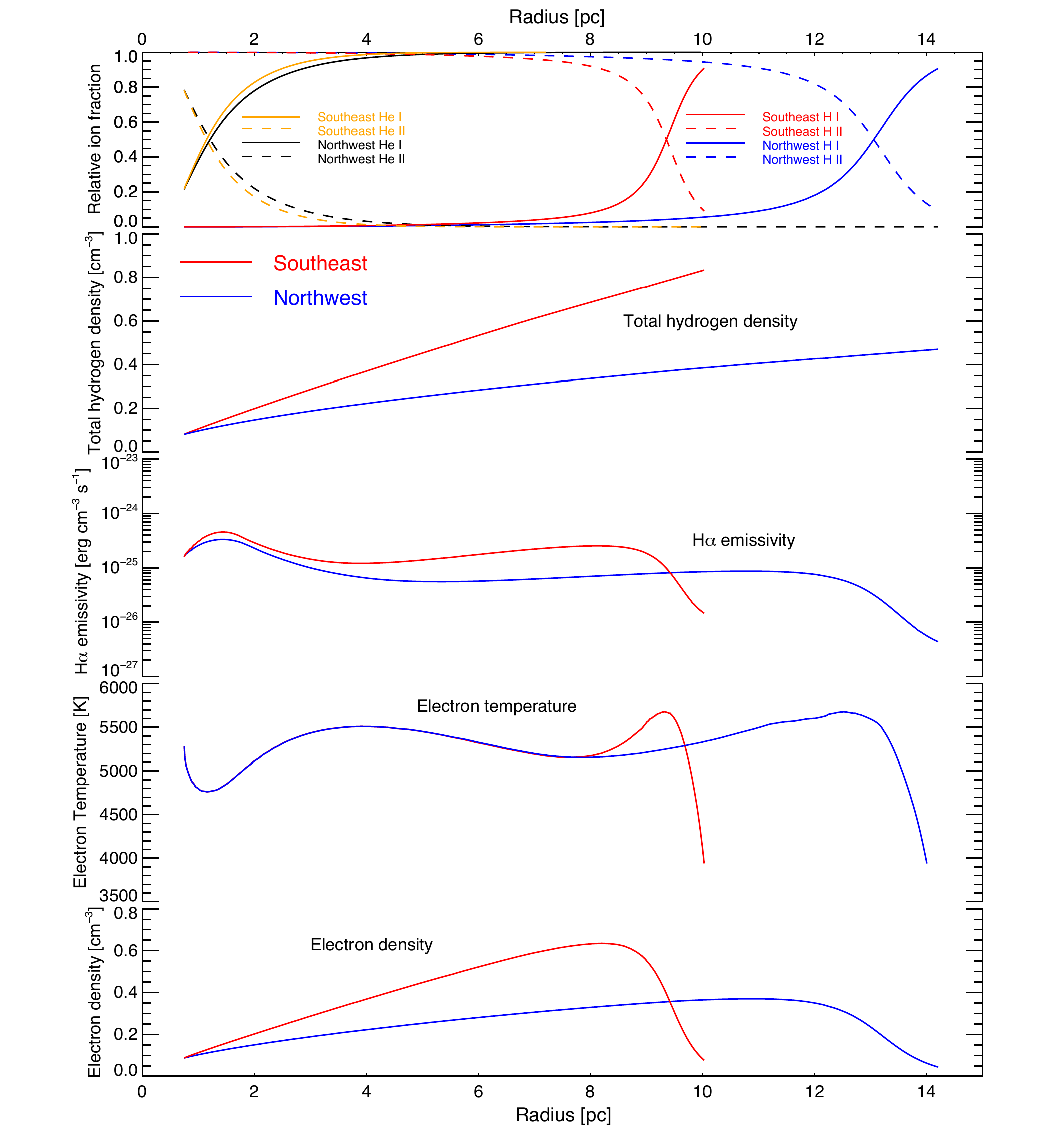}
    \caption{An example Cloudy 1D model structure for the southeast and northwest quadrants with a Lyman continuum $\log_{10} L_{\rm Ly}$ = 46.14: From the top: hydrogen and helium ionization structure, total hydrogen number density, \halpha\ volume emissivity, electron temperature and electron density.
    This matching pair of models has $\log_{10} r_0 = 18.36\, 
    {\rm cm}$, $n_0 = 0.08\, {\rm cm^{-3}}$ with $\alpha = 0.6$ for the northwest quadrant and $\alpha = 0.9$ for the southeast quadrant.}
    \label{fig:cloudy_structure}
\end{figure*}

\begin{figure*}
    \centering
    \includegraphics[width=\textwidth]{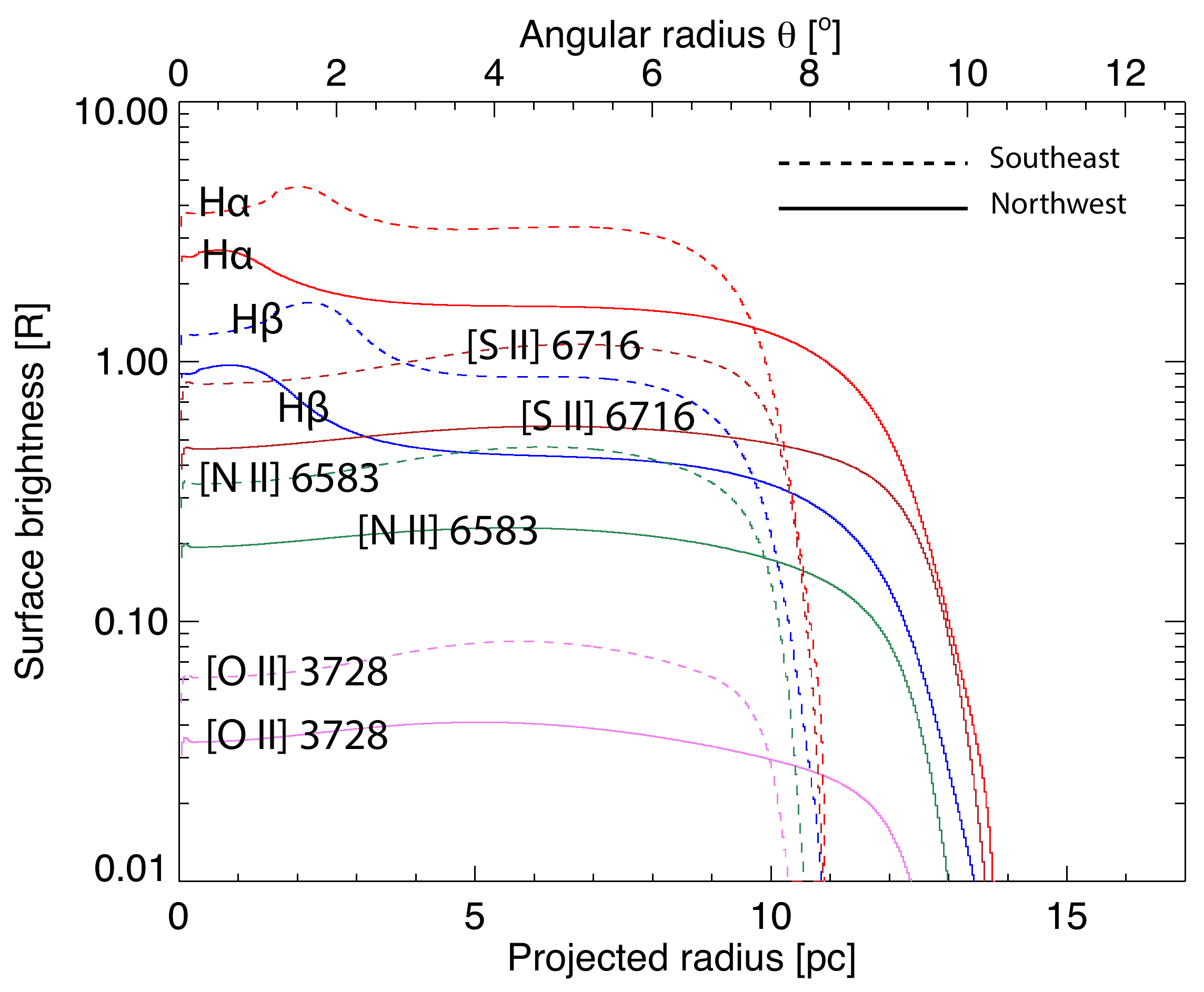}
    \caption{Synthetic surface brightness profiles 
of six emission lines for the northwest and southeast quadrants based on volume
emissivities from the same model pair show in Figure
\ref{fig:cloudy_structure}. 
    This matching pair of models has $\log_{10} r_0 = 18.36\, 
    {\rm cm}$, $n_0 = 0.08\, {\rm cm^{-3}}$ with $\alpha = 0.6$ for the northwest
    quadrant and $\alpha = 0.9$ for the southeast quadrant. 
    The models assume ISM abundances (by number) as defined by Cloudy version 17.02 (file: ism.abn): 
    S/H = 3.24$\times 10^{-5}$, N/H = 7.94$\times 10^{-5}$, O/H = 3.19$\times 10^{-4}$.}
    \label{fig:six_line_SB_profiles}
\end{figure*}

Figure \ref{fig:sb_profiles_for_matching_pairs} shows the synthetic
\halpha\ surface brightness profiles for the 352 model pairs along with
two models where the Lyman continuum varies with $\phi$ 
(see Figure \ref{fig:geometry} and Section \ref{sec:inclinations}) such
that different portions of the model spherical nebula see different projections
of Spica while in the direction of the Earth we see the $i=116^\circ$ view,
see Figure \ref{fig:three_inclinations_of_spica}a.   The model nebular 
structure for one pair of models is shown in Figure \ref{fig:cloudy_structure}.
Figure \ref{fig:six_line_SB_profiles}
shows an example of predicted surface brightness profiles for 5 lines in addition
to \halpha:
${\rm H\beta}$, {[\ion{S}{2}]} $\lambda$\,6716, {[\ion{N}{2}]} $\lambda$\,6583 and {[\ion{O}{2}]} $\lambda$\,3728.  An example Cloudy input script used to produce these data is
available on-line.

\section{Discussion} \label{sec:Discuss}
\subsection{Lyman continuum lower limit and leakage}
\label{discuss_lyman}
The lower limit on the integrated Lyman continuum established by  \cite{r85}, using
the revised Hipparcos distance of 76.6 pc  \citep{V07}, is $L_{\rm Ly} \ge 1.5\pm0.5\times 10^{46}\,
{\rm photons\ s^{-1}}$ or
$\log_{10} L_{\rm Ly} \ge 46.18^{+0.12}_{-0.18}$.
We find a consistent value for $L_{\rm Ly}$ when we sum all the SHASSA 
pixels $S(\alpha,\delta)$  within 8$^\circ$ of Spica, assuming 1 R of background and
incorporating the uncertainty in the distance $d=76.6\pm4.1$ pc:

\begin{equation}
L_{\rm Ly} \ge \frac{4\pi d^2}{\epsilon}\int S (\alpha,\delta)\,d\Omega = 1.59\pm 0.53\times 10^{46}\, {\rm photons\ s^{-1}}
 \label{eqn:lyman_lower_limit}
\end{equation}
or $\log_{10} L_{\rm Ly} \ge 46.22^{+0.13}_{-0.18}$ if, like R85, we assume   that any transitions in the Lyman series are optically thick within the nebula (Case B) 
with $\epsilon$ = 0.47, the number of \halpha\ photons per hydrogen recombination. R85 estimated $T$ = 8,000 K for the nebular 
electron temperature and cited \cite{pengelly64} for the value of $\epsilon$.
A value for $\epsilon$ is easily computed from

\begin{equation}
\epsilon = \frac{\alpha^{\rm eff}_{{\rm H}\alpha}}{\alpha_{\rm B}} = 
\frac{ \alpha^{\rm eff}_{{\rm H}\beta}}{\alpha_{\rm B}}
\frac{j_{{\rm H}\alpha}}{j_{{\rm H}\beta}}
\frac {\lambda_{{\rm H}\alpha}}{\lambda_{{\rm H}\beta}}
\end{equation}
where 
$\alpha^{\rm eff}_{{\rm H}\alpha}$  and
$\alpha^{\rm eff}_{{\rm H}\beta}$ are the effective recombination
rates for \halpha\ and \hbeta, respectively, 
$j_{{\rm H}\alpha}$ and $j_{{\rm H}\beta}$ are the
corresponding emission coefficients and 
${\alpha_{\rm B}}$ is the sum total
of all recombination rates in Case B.
The recombination rates and emission coefficients are temperature dependent \cite[see their Tables 2.1 and 4.2 and Equation 4.14]{agn3} and yield $\epsilon = 0.453$ at $T$=10,000 K
and $\epsilon = 0.487$ at $T$=5,000 K. Linearly interpolating to
T=8,000 K yields $\epsilon = 0.467$, consistent with the value
used by R85.  
The Cloudy models in Figure \ref{fig:cloudy_structure} have
a mean electron temperature close to 5,000 K, suggesting $\epsilon$
should slightly larger than the value used by R85, reducing
$\log_{10} L_{\rm Ly}$ by 0.03 dex.
The same Cloudy models have \halpha\ line  emissivities which are larger than Case B by 6\%, which corresponds
to a further 0.03 dex reduction.

The lower-lower limit, $\log_{10} L_{\rm Ly} \ge  46.04$ (calculated from
Equation \ref{eqn:lyman_lower_limit}),
is consistent with 46.04, the minimum value needed by 1D models
to match the median surface brightness profile for the southeast quadrant,
corresponding to a Lyman flux for $i = 130^\circ$, as shown in the distribution of inclinations in Figures \ref{fig:best_fit_north_south_independent} and 
\ref{fig:best_fit_north_south_match}.  
In a 3D model different portions of the nebula
will see Spica A from different vantage points, with 
$\log_{10} L_{\rm Ly}$ values up to 46.15 based on the stellar model for Spica A (see Table \ref{tab:lyman_fluxes}). The question of whether the orientation of the binary star within the nebula is fully consistent with the surface brightness distribution must await 3D photoionization models for the region.

\citet[R88]{r88} identified Spica as one of two stars 
where the hydrogen recombination rate of the surrounding nebula (based on the observed \halpha\ surface brightness) appeared to exceed the estimated Lyman continuum of the central star. 
R88 took the effective temperature of the second star, 139 Tau = HD 40111, spectral type B0.5II \citep{YBSC}, to be \teff = 20,400 K, while a more recent constraint is significantly
warmer: \teff = $25,771\pm 1684\, {\rm K}$ \citep{wu}.  The revision of the effective temperature upward for both stars resolves these mismatches identified by R88.

As another check on the ionization structure, R88 also measured the {[\ion{S}{2}]} $\lambda$\,6716/\halpha\ surface brightness ratio,
with a circular 50$\arcmin$ diameter field of view,
at two locations  in Spica's nebula near the north and west 
edges of the southeast quadrant: 
$0.16\pm0.03$ at (R.A., decl.) $\alpha = 201.3^\circ$, $\delta = -13.1^\circ$ ($2^\circ$ due south of Spica)
and $0.21\pm0.04$  at
$\alpha = 205.3^\circ$, $\delta = -11.1^\circ$ 
($3.5^\circ$ due east of Spica).  
The
{[\ion{S}{2}]} $\lambda$\,6716  to \halpha\ 
lines ratios 
 from the model profiles for the southeast quadrant
in Figure \ref{fig:six_line_SB_profiles}
yield ratios 30\% to 50\% higher than the observations: 0.22 at $2^\circ$ and 0.33 at $3.5^\circ$ from Spica. Our model assumes a S/H abundance ratio (by number) of 
$3.24\times 10^{-5}$ (Cloudy version 17.02 ISM abundances). 

\begin{figure*}[ht]
    \centering
    \includegraphics[width=\textwidth]{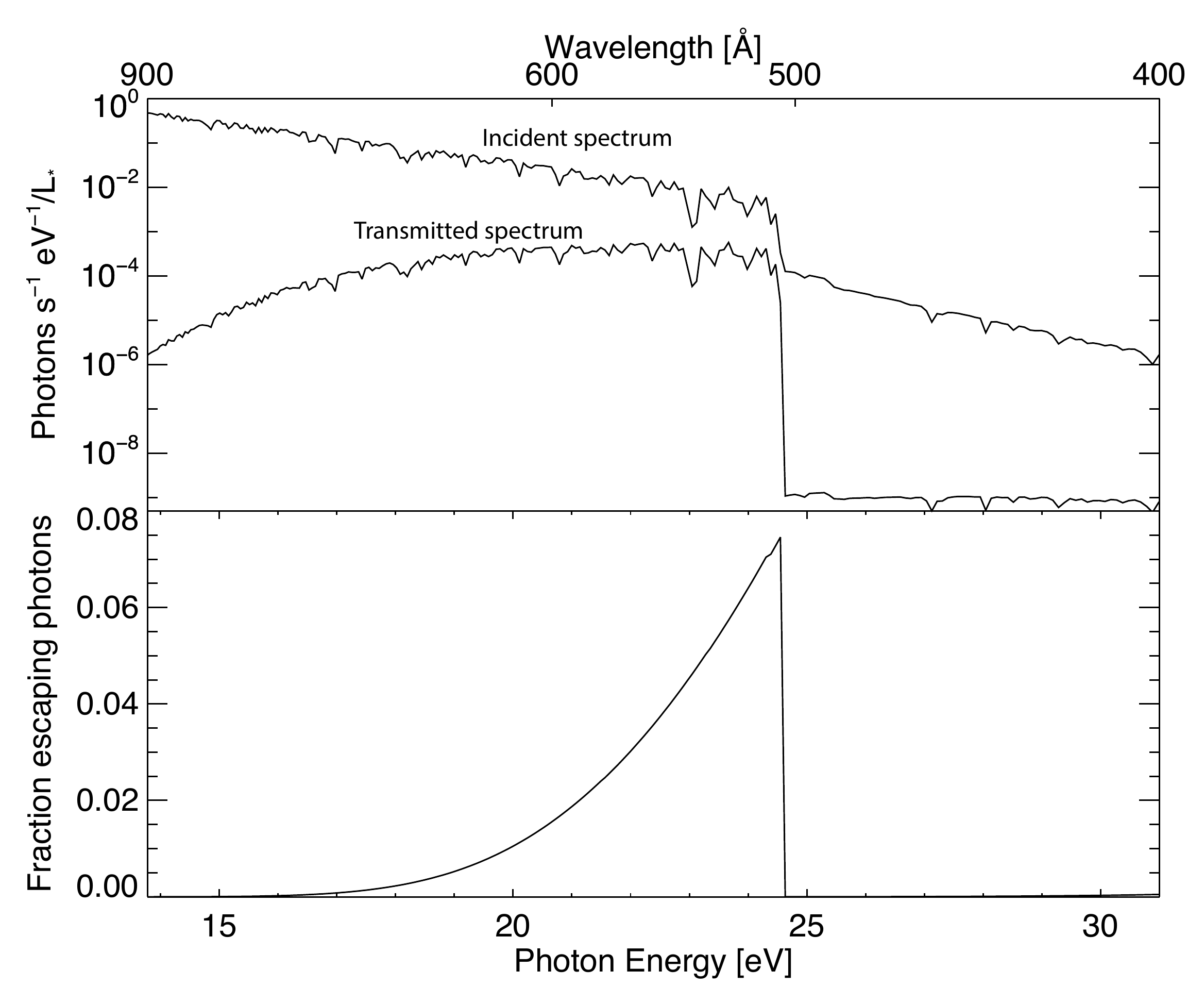}
    \caption{A model for the Lyman continuum escaping beyond 15 pc from 
   through the matter-bounded northwest portion of the nebula based
   on a 1D Cloudy model. Top panel shows the incident and 
   transmitted photon luminosity as a function of energy normalized 
    to the integrated Lyman 
    continuum value $L_{\rm Ly} = 1.38\times10^{46}\,{\rm photons\, s^{-1}}$,
    consistent with a view of Spica system with $i=170^\circ$ (see Table
    \ref{tab:lyman_fluxes}), and Cloudy parameters 
    $\log_{\rm 10} (r_0)=18.36\, \rm{cm}$, $n_0 = 0.08\, {\rm cm^{-3}} $ and $\alpha = 0.6$. For this model 16.8\% of
    the Lyman continuum photons escape, with over 7\% of photons escaping just
    below the bound-free edge of neutral helium at 24.6 eV.}
    \label{fig:lyman_leak}
\end{figure*}

While the southeast quadrant appears be radiation bounded by the
\ion{H}{1} cloud (see Figure \ref{fig:SHASSA_MAP}), the northwest quadrant
appears
to be matter bounded allowing some leakage of Spica's Lyman continuum into
the 
diffuse interstellar medium.  Figure \ref{fig:lyman_leak} shows that a model
for the transmitted Lyman
continuum (for northwest quadrant structure
in Figure \ref{fig:cloudy_structure})
is hardened between 13.6 eV and 24.6 eV and is effectively
extinguished
for energies above the ground-state bound-free edge of \ion{He}{1}, with 
 the integrated transmitted Lyman continuum being 17\% of the incident $L_{\rm Ly}$.
\citet[see their Figure 11d]{WM2004} found a similar effect for models tuned
to leak 15\% of the Lyman continuum 
around a much  hotter star (40,000 K).  The lack of a significant
\ion{He}{2} region surrounding Spica (helium is 100\% neutral beyond 4 pc, see Figure \ref{fig:cloudy_structure})
explains why the He-ionizing continuum is suppressed by more
than four orders of magnitude, compared to 
only one order of magnitude for the case investigated by \cite{WM2004}.

\subsection{Spica's far-UV absolute flux and line-of-sight extinction}
\label{discuss_EURD}
Our best fit model to the observed SED assumes the extinction curve from \cite{CCM}
and is significantly below the EURD flux level at wavelengths below 1070\,\AA\ 
(see Figure \ref{fig:spica_sed_comp}), although 
the residuals appear to be are uniformly offset,
apart from a strong residual deviation near Ly$\gamma$, suggestive of a
single systematic effect.
While the EURD spectrum has a very high S/N, more
than 3000 \citep{EURD}, the absolute flux calibration
uncertainty has been estimated to be $\pm20$\%  \citep{Bowyer1997,EURD_cal}, 
as reflected in the flux error bars in Figure \ref{fig:spica_sed_comp}. 
While \cite{EURD} states the absolute flux calibration is based on simultaneous
observations of the moon with EUVE, \cite{EURD_cal} states: ``We determined the EURD counts-to-flux conversion factor using... longward of 912 Å, to fits to stellar spectra."  This is consistent 
with  long-wavelength spectrometer
of EURD covering 500\,\AA\ to 1100\,\AA\ while the EUVE 
long-wavelength spectrometer reaches a maximum wavelength of 790\,\AA, 
leaving the 790\,\AA\ - 1100\,\AA\ range potentially subject to additional 
uncertainly from fits to unspecified stellar spectra.  On the other hand,
in support of the EURD absolute flux calibration are Voyager UVS observations
\citep{Voyager} that are in good agreement \citep{EURD}.

\begin{figure*}
    \centering
    \includegraphics[width=\textwidth]{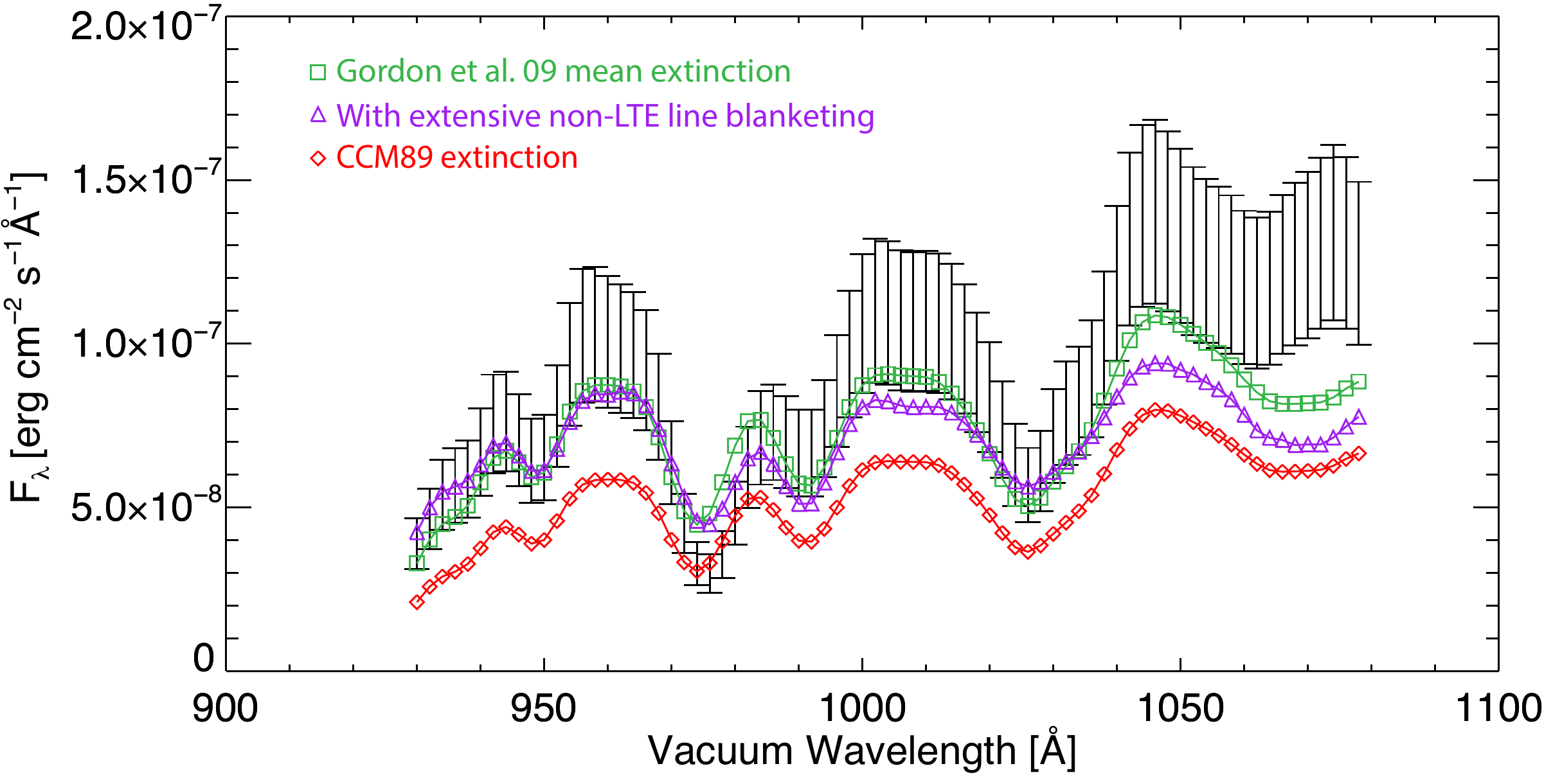}
    \caption{The EURD spectrum from \cite{EURD} binned to 2\,\AA\ with
    $\pm 20\%$ error bars compared to synthetic spectra convolved with
    a Gaussian to $R=\lambda/{\Delta\lambda} = 200$, the resolution of EURD, and binned to 2\,\AA.  The lowest (red) model spectrum is the
    same one shown in Figure \ref{fig:spica_sed_comp} with the extinction
    curve of \cite{CCM} parameterized by the color excess, $E(B-V) =
    0.0275$, and ratio of total of selective to visual extinction, $R(V)
    = 3.039$. The upper (green) spectrum uses the extinction
    parameterization of \cite{FM90} with the mean parameters from
    \cite{Gordon09}: $C_1^{A(V)} = 0.8161, C_2^{A(V)} = 0.2559,
    C_3^{A(V)} = 1.0915, C_4^{A(V)} = 0.1142, x_0 = 4.5909,\gamma =
    1.0076$, with same $E(B-V)$ and $R(V)$ values. The middle (purple)
    spectrum has the same extinction as the upper (green) spectrum but is 
    single model atmosphere parameterized  by
    $T_{\rm eff}  = 24\,500\,{\rm K}, \log_{10} (g) = 3.35$, R = 9.19
    $\nom{R}$, solar abundances \citep{A09}, and non-LTE species
    \ion{H}{1}, \ion{He}{1-II}, \ion{C}{1-IV}, \ion{N}{1-IV}, \ion{O}{1-V},
    \ion{Mg}{1-IV}, \ion{Al}{1-V},    \ion{Si}{1-V},
    \ion{S}{1-IV}, \ion{Sc}{1-IV}, \ion{Ti}{1-V}, 
    \ion{Mn}{1-IV}, \ion{Fe}{1-V}, \ion{Co}{1-V}, \ion{Ni}{1-V} with other
    species in LTE. Alignment of the principal features is good shortward of
    1050\,\AA. With extensive non-LTE line-blanketing the alignment of the Lyman $\gamma$ feature near 975\,\AA\ is improved.}
    \label{fig:EURD_closeup}
\end{figure*}

The EURD absolute flux level raises questions about
best mean temperature for Spica A and far-UV extinction curve towards Spica.
It may be possible for a cooler stellar model and weaker far-UV extinction 
to provide a better fit overall, however our work suggests a cooler stellar model will not
have a Lyman continuum sufficient to produce the observed \halpha\ surface brightness
profiles, particularly in the southeast quadrant where our models indicate $L_{\rm Ly} \ge
1.1 \times 10^{46}$.  A possible solution is an improved extinction curve. The analysis by
\cite{Gordon09}, using both IUE and FUSE extinction curves, found the \cite{FM90} parameter $C_4^{A(V)}$, which describes the far-UV rise,
to be 8\% weaker on average, while other extinction parameters were
consistent with previous IUE-only analyses.  Adopting the mean
extinction parameters \cite[see Table 3 therein]{Gordon09},
the model comparison to the EURD spectrum is improved, as shown in 
Figure \ref{fig:EURD_closeup}, because the far-UV extinction is reduced for the same $E(B-V)$ and $R_V$ values.  If the far-UV extinction curve towards
Spica is similar or weaker than the \cite{Gordon09} mean values, then the tension
between the best stellar models for the full SED, including the EURD spectrum, and the
best stellar models to produce the \halpha\ emission would be reduced.  While beyond the
scope of the present study, a simultaneous fit including all of  Spica's parameters and all six of the \cite{FM90} extinction parameters could better
characterize the extinction curve and model atmosphere.

\subsection{Comparisons to an earlier analysis of the SHASSA data}
\label{discuss_park}
The analysis by \citet[P2010]{Park2010}
studied general trends in \halpha\ surface brightness profiles
for the northern and southern portions of the nebula.
They found an inner cloud radius of $r_0=0.3$ pc and a constant hydrogen number density, $n=0.22\, {\rm cm^{-3}}$, for the northern half of the nebula 
and an increasing density profile, starting with $n_0=0.22\, {\rm cm^{-3}}$
and a power-law index $\alpha = 0.15$ (see Equation \ref{eqn:density_law}), for the southern half
of the nebula. 

Our analysis indicates a steeper density increase than P2010:
$0.3 \le \alpha \le 0.9$ for the northwest quadrant and
$0.8 \le \alpha \le 1.2$ for the southeast quadrant.
Our initial density range $0.04 \le n_0 \le 0.20 {\rm\, cm^{-3}}$
and inner radius range $0.1 \le r_0 \le 1.7 {\rm\,  pc}$
are tightly positively correlated (see Figure \ref{fig:best_fit_north_south_match}).

P2010 choose an effective temperature
for Spica A of 26,000 K which is significantly
hotter than our mean effective temperature of 24,777 K for Spica A.
At its apparently most luminous orientation (pole-on), our model for Spica A has a  polar temperature of 25,642 K, $\log_{10}(L_{\rm Ly})$ = 46.15 (see Table \ref{tab:lyman_fluxes}), which 12\% less than the value used by P2010, 46.2.

Our nebular parameters  differ from P2010
in part because we are comparing models to different surface brightness profiles.
P2010 used nine evenly-spaced $1^\circ$-wide annular bins
and smoothed the SHASSA data to $3^\circ$,
compared to our much narrower 60 bins of equal area
with no smoothing (see Figure \ref{fig:SB_profiles_park_comparison}).
Our profile shapes differ most from P2010 within $0.5^\circ$ of the star.
Although from the same SHASSA data set, the P2010
\halpha\ surface brightness values are $\sim\,70\%$ larger in this inner region.
P2010 could possibly have employed the SHASSA
stellar mask differently allowing more pixels to be contaminated with stellar continuum emission, contributing to an apparently higher \halpha\ surface brightness.
We were unable to reproduce the innermost P2010 surface brightness 
values by neglecting to use the stellar mask, so the cause of this
discrepancy remains uncertain. Overall we found the use of the stellar mask
reduces the surface brightness variance within each bin relative to P2010. 
Our narrower annuli better sample the abrupt drop 
in surface brightness starting $\sim\,5.5^\circ$ from Spica in the southeast
quadrant of the nebula. As a result, 
we found a higher $\alpha$ value, 
hence a  steeper density rise is required match the data as the 
total hydrogen density ramps up toward the neutral cloud to the southeast, see Figure \ref{fig:SHASSA_MAP}.

\subsection{Inner cloud radius and Spica's Mass-loss Rate}
\label{discuss_wind}
We can compare our range of inner cloud radii, $r_0$ values, to the standoff distance, $R_0$, in shock nebulae around low proper motion stars 
in the recent work by \citet[K2019]{ChipK2019}.
Low proper motion for these stars suggests that the shock is primarily wind-driven rather than being a motional shock.  Spica A's nominal spectral type is B1 III-IV \citep{YBSC} and two stars from K2019, HD 240015 (B0 III) and HD 240016 (B2 III), have $R_0$ values of 0.215 pc and 0.174 pc,
with corresponding ambient interstellar number densities of 
5\,cm$^{-3}$ and 8\,cm$^{-3}$ respectively.  While these radii fall within
our estimates $0.1 \le r_0 \le 1.7 {\rm\,  pc}$, 
the densities are about 100 times larger than our range
of $n_0$ values. 

We can estimate Spica A's mass-loss rate 
following \citet[Equation 3 therein]{ChipK2018}, 
\begin{equation}
    \dot{M}=\frac{4\pi R_0^2V_a^2 n_a m_H}{V_w},
    \label{eqn:mdot_k2018}
\end{equation}
which balances the momentum flux between the stellar
wind and the impinging interstellar material,
where $V_a$ is the motion of the star relative to the
ambient impinging medium,
$n_a$ is the number density of the ambient medium, $m_H$ is
the hydrogen mass
and $V_w$ is the stellar wind speed. Spica's center-of-mass space motion
is dominated by the transverse component (see proper motion vector in 
Figure \ref{fig:SHASSA_MAP}) which yields $V_a = 19\,\rm km\,s^{-1}$
at a distance of 76.6 pc. The radial velocity of the center of mass, $\gamma$, is less than two kilometers per second \citep{MOST_2016}.
A lower limit on the wind speed is the escape velocity from the hot pole of Spica A, 
\begin{equation}
V_w \ge   \sqrt{\frac{2GM_A}{R_{\rm pole}}} = 735\,\rm km\,s^{-1},
\end{equation}
where we adopt $M_A=10.0\,\nom{M}$ and $R_{\rm pole} = 7.06\,\nom{R}$, consistent
with the stellar parameters in Figure \ref{fig:three_inclinations_of_spica}.
Adopting $R_0=r_0=0.74$\, pc and $n_a = n_0=0.08\,\rm cm^{-3}$ (near the mode
of the distributions of $r_0$ and $n_0$ in  Figure \ref{fig:sb_profiles_for_matching_pairs}),
consistent with model nebular structure shown in Figure \ref{fig:cloudy_structure}, 
yields $71\times 10^{-10}\, M_\odot\,\rm yr^{-1}$, a mass-loss rate
with a similar order of magnitude to estimates from K2019: HD 240015 (B0 III) 
with $22\times 10^{-10}\, M_\odot\,\rm yr^{-1}$ 
and HD 240016 (B2 III) with $245\times 10^{-10}\, M_\odot\,\rm yr^{-1}$.
We can also compare these values
to the mass-loss predicted for Spica A from the theoretical modified wind
momentum-luminosity relationship of \citet[Equation 1 therein]{KK2007}: $140 \times 10^{-10}\,
M_\odot\,\rm yr^{-1}$
for $\log_{10}L/\nom{L} = 4.26$, $\bar{R}/\nom{R} = 7.37$ and $v_\infty = V_w = 735\, {\rm
km\,s^{-1}}$.  This value is within
a factor of two of that derived from Equation \ref{eqn:mdot_k2018}. If the stellar wind
from Spica A creates
a visible shock at $r=0.74$\, pc, 3.3$\arcmin$ from the star, it is not
resolved in the SHASSA data set with an angular resolution of 6$\arcmin$.

\subsection{Comparisons to line-of-sight column densities}
\label{discuss_column_densities}
The analysis by \citet[YK79]{york_kinahan}, 
of ultraviolet line absorption along the line of sight towards Spica, 
estimated half of the column through the local hydrogen to be in the \ion{H}{2} region, the other half being neutral gas:
$N(\text{H}) = N(\text{\ion{H}{2}}) = 1.0 \pm 0.1 \times 
10^{19}\,{\rm cm^{-2}}$.  Integrating the electron density, $n_e$,
profiles in
Figure \ref{fig:cloudy_structure},
\begin{equation}
    \label{eq:n_e_column}
    N(\text{\ion{H}{2}})\simeq N(e^-) =\int_0^{R_n} n_e(r)\, dr
\end{equation}
(where $R_n\equiv r_0+\mathrm{maximum}\ \mathrm{depth}$)
yields $1.09\times 10^{19}\ \rm{cm}^{-2}$ in the northwest and
$1.16\times 10^{19}\ \rm{cm}^{-2}$ in the southeast.
This comparison supports the YK79's assumption of a single \ion{H}{2}
region along the line of sight as the model accounts for the total
estimated electron column density. 

Using the ratios of column densities for lines that probe
the excitation of the ground term fine structure of \ion{N}{2},
YK79 constrained the electron density 
in the \ion{H}{2} region to be $n_e \leq 0.5\, {\rm cm^{-3}}$, quite
consistent with the model electron-density values in Figure
\ref{fig:cloudy_structure}.
Adopting $n_e = 0.5\,{\rm cm^{-3}}$,  
YK79 deduced an \ion{H}{2} region radius  
$R_n = N(\text{\ion{H}{2}})/n_e$ = 6.7 pc. 
Our radii are larger, 10 to 14 pc, 
consistent with projected dimensions of the 
\halpha\ nebula on the sky, with
corresponding lower mean electron densities 
for the same column density.
Dividing Equation \ref{eq:n_e_column}
by the $R_n$ values, corresponding to the maximum radii in
Figure \ref{fig:cloudy_structure}, 
yields mean electron density values
$\bar{n}_e=0.25$\,cm$^{-3}$ for the northwest quadrant
and $\bar{n}_e=0.37$\,cm$^{-3}$ for the southeast quadrant.

Model column densities through the \ion{H}{2} region
are compared to the values from YK79 in Table \ref{tab:column_densities}.  
For a realistic model, the \ion{H}{2} region column densities must be equal to or less than
the total column density along the line of sight to Spica, particularly for 
neutral species which could be present outside of the ionized gas 
surrounding Spica.  Model column densities for
two neutral species, \ion{C}{1} and \ion{Mg}{1}, are significantly
larger than YK79's values by $\simeq +0.3$ dex, however the oscillator
strengths for the lines employed by YK79 are now lower by $-0.24$ dex
and $-0.16$ dex \citep{NIST}, respectively, which increases the inferred YK79 column
densities by the same amount.  
For other neutral species, the model indicates $\sim$40\% of \ion{N}{1} 
resides in the \ion{H}{2} region,  $\sim$20\% for \ion{O}{1},
with a column density for \ion{S}{1} consistent with YK79's upper limit.
For the ionized species,  20\% to 50\%
of \ion{C}{2} is predicted to be in the \ion{H}{2} region,
50\% to 95\% of \ion{Mg}{2} and essentially all of the
\ion{S}{2} and \ion{S}{3} ions.

\begin{deluxetable*}{lCCCC}
\tablecaption{Model \ion{H}{2} region column densities compared to YK79 values with
oscillator strength update}
\label{tab:column_densities}
\tablehead{\colhead{Ion} &\colhead{Model $\log_{10}N$} &\colhead{YK79 $\log_{10}N$}
&\colhead{$\Delta\log_{10} N$} &\colhead{Max $\Delta \log_{10} f$}\\ 
\colhead{} &\colhead{(cm$^{-2}$)} &\colhead{(cm$^{-2}$)}
&\colhead{Model - YK79} &\colhead{NIST - YK79}}
\decimals
\startdata
\ion{C}{1}  &11.95 &11.66^{+0.21}_{-0.46} &$+0.29^{+0.46}_{-0.21}$ &$-0.24$\tablenotemark{a}\\
\ion{C}{2}  &15.53 &16.00\pm0.2  &$-0.47\pm0.2$ &$+0.04$\tablenotemark{b}\\ 
\ion{N}{1}  &14.23 &14.66^{+0.04}_{-0.06} &$-0.43^{+0.06}_{-0.04}$ &$+0.04$\tablenotemark{c} \\
\ion{N}{2}  &14.92 &14.40\leq\log_{10}{N}\leq16.12  &\nodata &\nodata\\
\ion{O}{1}  &14.87 &15.58\pm0.10  &$-0.71\pm0.1$ &$-0.13$\tablenotemark{d} \\
\ion{Mg}{1} &14.08 &13.8\pm0.1 &$+0.28\pm0.1$ &$-0.16$\tablenotemark{e} \\
\ion{Mg}{2} &11.38 &11.5^{+0.2}_{-0.1}  &$-0.12^{+0.1}_{-0.2}$ &$-0.20$\tablenotemark{f} \\
\ion{S}{1}  &10.53 &< 11.49  &\nodata &\nodata \\
\ion{S}{2}  &14.60 &14.51\pm0.04  &$+0.1\pm0.04$ &$+0.06$\tablenotemark{g} \\
\ion{S}{3}  &13.51 &13.56\pm0.04  &$-0.05\pm0.04$ &$+0.09$\tablenotemark{h} \\
\enddata
\tablecomments{Model column density, $\log_{10}N$, values calculated from a Cloudy model with 
$\log_{10} L_{\rm Ly}=46.14$, $\log_{10} r_0 = 18.36\, 
{\rm cm}$, $n_0 = 0.08\, {\rm cm^{-3}}$, $\alpha = 0.9$, see Figure \ref{fig:cloudy_structure},
with abundances C/H = $2.51\times 10^{-4}$,
N/H = $7.94\times 10^{-5}$, O/H=$3.19\times 10^{-4}$, Mg/H = 1.26$\times10^{-5}$
and S/H = $3.24\times 10^{-5}$. Updated oscillator strengths, $f$, are from \cite{NIST}.}
\tablenotetext{a}{\ion{C}{1} 1277.285\,\AA: $-1.04$  (NIST)  $-0.80$  (YK79)}
\tablenotetext{b}{\ion{C}{2} 1334.542\,\AA: $-0.89$  (NIST)  $-0.93$  (YK79)}
\tablenotetext{c}{\ion{N}{1} 1134.415\,\AA: $-1.84$  (NIST)  $-1.87$  (YK79)}
\tablenotetext{d}{\ion{O}{1} 1039.230\,\AA: $-2.04$  (NIST)  $-1.91$  (YK79)}
\tablenotetext{e}{\ion{Mg}{1} 2025.824\,\AA: $-0.95$  (NIST)  $-0.79$  (YK79)}
\tablenotetext{f}{\ion{Mg}{2} 1239.936\,\AA: $-3.21$  (NIST)  $-3.01$  (YK79)}
\tablenotetext{g}{\ion{S}{2} 1259.518\,\AA: $-1.74$  (NIST)  $-1.80$  (YK79)}
\tablenotetext{h}{\ion{S}{3} 1012.504\,\AA: $-1.36$  (NIST)  $-1.45$  (YK79)}
\end{deluxetable*}

\section{Summary} 

\begin{enumerate}
\item Our analysis of the \halpha\ emission surrounding Spica
has incorporated for the first time a model for the spectral energy distribution (SED) of the central stars 
which is shown to match archival absolute spectrophotometry of Spica between 930\,\AA\ and  10200\,\AA\ (see Section \ref{sec:spica_model} and Figure \ref{fig:spica_sed_comp} and \ref{fig:EURD_closeup}).
Our estimate for the primary star mean effective temperature, $\simeq$\,24,800 K, is significantly warmer 
($+3300$\,K) than used in earlier work \citep{r85,r88} which found 
the predicted Lyman continuum luminosity, $L_{\rm Ly}$, from the contemporary model atmosphere was insufficient to produce
the measured hydrogen recombination rate determined from \halpha\ surface brightness measurements.   We find a lower limit of $\log_{10} L_{\rm Ly} \ge 46.22^{+0.13}_{-0.18}$
by integrating over the SHASSA map within $8^\circ$ of Spica. This value is consistent with
R85's value at a distance 76.6 pc, $\log_{10} L_{\rm Ly} \ge 46.18^{+0.12}_{-0.18}$.

\item Our model SED for Spica is significantly below the measured
absolute flux in the far-UV  from 930\,\AA\ to 1080\,\AA\ \citep{EURD} if we use the extinction curve from \cite{CCM} (see Figure \ref{fig:EURD_closeup} and Section \ref{discuss_EURD}).  The updated mean
extinction curve from \cite{Gordon09}, based in part on FUSE observations,
provides a closer match with a color excess, $E(B-V)=0.0275$, consistent with earlier estimates, e.g. 0.03 from \cite{YBSC}.

\item In this work the primary star was modeled as a tidally-distorted, rapid rotator (see 
Figure \ref{fig:three_inclinations_of_spica},
Section \ref{sec:spica_model} and
Appendix \ref{sec:appendixA}). 
The poles of the star are about 900\,K hotter than the mean effective temperature,
which results in a factor of 1.6 difference in the apparent $L_{\rm Ly}$
between the pole-on view and equator-on view (see Table \ref{tab:lyman_fluxes}), similar to the
factor of two change in $L_{\rm Ly}$ for single-temperature models \citep[see their Table 4]{LH2007}.
with $\Delta$\teff =1000\,K  at 25,000 K.

\item Using the SEDs from this stellar model
in various orientations (see Section \ref{sec:spica_model} and
Table \ref{tab:SED_data}) as input to 1D photoionization models, and generating synthetic
\halpha\ surface brightness profiles 
(see Section \ref{sec:comp_sb_section})
for comparison to observations
(see Figure \ref{fig:SB_profiles_park_comparison} and Section \ref{sec:extraction}), indicates
that $\log_{10} L_{\rm Ly} > 45.99$ in the northwest quadrant and 
and $\log_{10} L_{\rm Ly} > 46.04$ in the southeast quadrant
of the \ion{H}{2} region consistent with \cite{r85,r88} with the updated distance of 76.6 pc (see Section \ref{discuss_lyman}). For the matter-bounded northwest quadrant (see Figure \ref{fig:SHASSA_MAP}),
models for the transmitted spectrum predict about 17\% of $L_{\rm Ly}$ leaks from the
\ion{H}{2} region with the helium-ionizing continuum suppressed by four orders of
magnitude (see Figure \ref{fig:lyman_leak}).

\item After comparing nearly 1.5 million Cloudy photoionization models to median \halpha\ surface brightness profiles from the southeast and northwest quadrants of the \ion{H}{2} region (see Section \ref{sec:results}), we identified
352 pairs of models which provide a reasonable match
to the data (see Figure \ref{fig:sb_profiles_for_matching_pairs}). These models have the total hydrogen density
increasing away from the star, starting from an inner boundary,
$r_0 \simeq 0.8^{+0.3}_{-0.4}$ pc, and initial number density 
$n_0 \simeq 0.07^{+0.04}_{-0.03}\,{\rm cm^{-3}}$, which are tightly correlated (see Figure \ref{fig:best_fit_north_south_match}). 
The density as a function of radius is parameterized as a power law with an index  $\alpha \simeq 1.0\pm0.1$ in the southeast quadrant
and $\alpha \simeq 0.7^{+0.1}_{-0.2}$ in the northwest quadrant.
These indexes are significantly higher than
reported by \cite{Park2010} who found a constant density profile
($\alpha = 0$) for the northern nebula and $\alpha=0.15$ for
the southern nebula (see Section \ref{discuss_park}) though our mean total hydrogen densities, 
near $0.3\, {\rm cm^{-3}}$, are quite similar (see Figure \ref{fig:cloudy_structure}).

\item The validity of the nebular structure (see Figure \ref{fig:cloudy_structure})
could be further tested by surface brightness measurements in other lines,
for example: ${\rm H\beta}$, {[\ion{S}{2}]} $\lambda$\,6716, {[\ion{N}{2}]} $\lambda$\,6583 and {[\ion{O}{2}]} $\lambda$\,3728 (see Figure \ref{fig:six_line_SB_profiles}).
The model {[\ion{S}{2}]} $\lambda$\,6716-to-\halpha\ ratios
are 30\% to 50\% larger than two observations 
by \cite{r88} east and south of Spica (see Section \ref{discuss_lyman}). The limitations of 1D models
for Spica's nebula should be explored with 3D models,
for example \citet{Wood2013}.

\item An inner boundary radius of 0.74 pc is
consistent with wind-driven shock for a stellar mass-loss rate
$\sim 10^{-8}\,M_\odot\,\rm yr^{-1}$ and a terminal velocity exceeding $735\,{\rm km\, s^{-1}}$ (see Section \ref{discuss_wind}). The shock,
if visible, would be a few arc minutes from the star.

\item Model \ion{H}{2} region column densities for 
\ion{C}{1},
\ion{C}{2},
\ion{N}{1},
\ion{N}{2},
\ion{O}{1}, 
\ion{Mg}{1},
\ion{Mg}{2},
\ion{S}{1},
\ion{S}{2} and \ion{S}{3}
are largely consistent with
the total line-of-sight column densities from \cite{york_kinahan}
after updating their oscillator strengths for \ion{C}{1} and \ion{Mg}{1}
(see Section \ref{discuss_column_densities}).

\end{enumerate}


\acknowledgments
This work made use of the Cray CS400 
High Performance Cluster ``Vega'' at Embry-Riddle Aeronautical University (ERAU).

Support was provided by the Nadine Barlow Undergraduate Research Support Award
from the Physics and Astronomy  division of the  
Council on Undergraduate Research and ERAU's Office of Undergraduate Research. 
The EURD and IUE spectra archives have been developed in the framework of the Spanish Virtual Observatory project supported by the Spanish MINECO/FEDER through grant AyA2017-84089. The system is maintained by the Data Archive Unit of the CAB (CSIC -INTA). Thank you to Enrique Solano Marquez for providing access to the EURD data set. This research has made use of the VizieR catalogue access tool, CDS, Strasbourg, France (DOI: 10.26093/cds/vizier). The original description of the VizieR service was published in A\&AS 143, 23. We thank the anonymous referee for the careful reading, comments and questions that improved the manuscript.
Thanks to Peter Hauschildt and Travis Barman for advice on computing
high-resolution synthetic spectra with PHOENIX.  Thanks to Matt Haffner for comments on the manuscript.
Thanks to Gary Ferland and the Cloudy team at the University of Kentucky for the Cloudy workshops they host.
J. Hammill was trained at the Summer 2019 workshop in Lexington, KY.

%

\vspace{5mm}
\software{PHOENIX (v18.02.00A; Hauschildt \& Baron 1999, 2010, 2014), Cloudy (v17.02; Ferland et al. 2017), corner.py (Foreman-Mackey 2016)}





\appendix
\section{Binary Star Model Atmosphere Construction}
\label{sec:appendixA}
Our code, written in the Interactive Data Language (IDL), follows the outline
of \cite{wilson79} and computes the surfaces of binary stars in close,
eccentric orbits from an approximate gravitational potential and
interpolates model atmosphere intensities onto these surfaces. Furthermore
the code computes how these surfaces are then projected onto the plane of the sky, from which a synthetic intensity maps and spectral energy distributions
are computed.

\subsection{Coordinate System and Potentials}
For circular orbits with synchronous rotating
components, the surface forms of the stellar components are defined
by equipotential surfaces of centrally-condensed point masses (Roche
Models).  For the case of eccentric orbits like Spica, the distance
between the centers of the components changes with orbital phase, so the
equipotential surfaces must be computed at each phase point under the
assumption that each component readjusts to equilibrium on time scales
much faster than the orbital period.  Uniformly rotating yet
asynchronous stellar components have no consistent simple solution.  To
establish constant-density surfaces for the components, we start
with the approximate potential 
following \citet[Equation 16 therein]{limber63}
understanding that it is not strictly valid due to the neglect of
small Coriolis forces from stellar fluid flow.  This potential assumes
that both rotational angular velocity vectors and the orbital angular
velocity vector are parallel.

The potential describing the surface of the primary in rectangular coordinates is

\begin{equation}
V_1 = - G\frac{m_1}{r_1} - G\frac{m_2}{r_2} - \frac{1}{2}\omega^2_{\rm orb}
\left[\left(\frac{\omega_{\rm 1\ rot}}{\omega_{\rm orb}}\right)^2 - 1\right](x^2+y^2)
-\frac{1}{2}\omega^2_{\rm orb}\left[\left(x - \frac{D m_2}{m_1+m_2}\right)^2 + y^2\right],
\label{eqn:v1}
\end{equation}
where the origin of the coordinate system is at the center of primary, in
the frame which rotates about an axis perpendicular to the orbital
plane at a constant rate with the period of the binary, $P$.  The x-axis
is in the orbital plane along the line of centers.
The first term in Equation \ref{eqn:v1} represents the potential from the primary where $G$ is
the gravitational constant, $m_1$ the mass of the primary, and
\begin{equation}
r_1 = (x^2 + y^2 + z^2)^{1/2}
\end{equation}
is the distance from the center of mass of the primary.  The second term 
represents the potential from the secondary mass $m_2$  where
\begin{equation}
r_2 = \bigl((D-x)^2 + y^2 + z^2\bigr)^{1/2}
\end{equation}
is the distance from the center of mass of the secondary,  $D$ is
the instantaneous separation of the centers of mass of the two stars.
In units of the semi-major axis, $D$ is computed from
the eccentricity $e$ and the true anomaly $\upsilon$:

\begin{equation}
D = \frac{1-e^2}{1+e\cos\upsilon} \label{eqn:seperation}. 
\end{equation}

The coordinates ($x,y,z$) of the primary point mass 
are (0,0,0) and for secondary mass ($D,0,0$).  The third term in Equation \ref{eqn:v1} represents the contribution from
the asynchronous rotation where $\omega_{\rm orb}$ is the angular
orbital rate ($2\pi/P$) and $\omega_{\rm 1\ rot}$ is the angular rotational rate of the
primary. This term vanishes for synchronicity: $\omega_{\rm 1\ rot} = \omega_{\rm
orb}$.  The fourth term represents potential from the centrifugal
acceleration due to the rotating frame.

For this asynchronous system a separate potential is needed to describe
the surface of the secondary in rectangular coordinates:
\begin{equation}
V_2 = - G\frac{m_1}{r_1} - G\frac{m_2}{r_2} - \frac{\omega^2_{\rm orb}}{2}
\left[\left(\frac{\omega_{\rm 2\ rot}}{\omega_{\rm orb}}\right)^2 - 1)\right]\left((D-x)^2 + y^2\right)
-\frac{1}{2}\omega^2_{\rm orb}\left[\left(x - \frac{D m_2}{m_1+m_2}\right)^2 + y^2\right]
\label{eqn:v2}
\end{equation}
Both potentials (Equations \ref{eqn:v1} and \ref{eqn:v2}) can be
greatly simplified when put in terms of the mass ratio $q = m_2/m_1$,
instantaneous separation $D$, the respective radius vectors $r_1$ and
$r_2$, the direction cosines of each radius vector,
\begin{eqnarray}
\lambda &= &\cos\varphi\sin\vartheta \label{eqn:lambda_cosine_radius_vector}\\
\mu &= &\sin\varphi\sin\vartheta\\
\nu &= &\cos\vartheta 
\end{eqnarray}
where $\varphi$ is the azimuthal angle (longitude) and $\vartheta$ is the polar angle (co-latitude).  The
rectangular coordinates are related to the spherical coordinates
for the primary star by
\begin{eqnarray}
x &=& r_1\ \lambda \\
y &=& r_1\ \mu \\
z &=& r_1\ \nu \\ \label{eqn:z_coord} 
\end{eqnarray}
and for the secondary star by
\begin{eqnarray}
x &=& D + r_2\ \lambda\\ 
y &=& r_2\ \mu \\
z &=& r_2\ \nu. 
\end{eqnarray}

The simplified potentials,
\begin{eqnarray}
\Omega_1 &=& \frac{-V_1}{Gm_1} -\frac{1}{2D}\left(\frac{q^2}{1+q}\right)\\
\Omega_2 &=& \frac{-V_2}{Gm_1} -\frac{1}{2D}\left(\frac{q^2}{1+q}\right)
\end{eqnarray}
for the primary and secondary stars respectively, differ from the original potentials $V_1$ and
$V_2$ by constants. The mass ratio $q$ is fixed for the model and the
instantaneous separation $D$ (Equation \ref{eqn:seperation}) is fixed
at a given epoch.
With the semi-major axis set to unity, Kepler's third law becomes,
\begin{equation}
\omega_{\rm orb}^2 = Gm_1 (1 + q),
\end{equation}
and the simplified potentials in terms of $q$, $D$, $r_1$, $r_2$, and the
direction cosines are


\begin{equation}
\Omega_1 = \frac{1}{r_1} + \frac{q}{\sqrt{D^2 + r_1^2 - 2r_1\lambda D}} - {qr_1\lambda D} + 
\frac{1}{2}\left(\frac{\omega_{\rm 1\ rot}}{\omega_{\rm orb}}\right)^2(1+q){r_1^2}(1-\nu^2) \label{eqn:omega1} \\
\end{equation}
\begin{equation}
\Omega_2 = \frac{1}{\sqrt{D^2 + r_2^2 + 2r_2\lambda D}} + \frac{q}{r_2} + \frac{1}{2}(1+q)(D^2 + 2Dr_2\lambda) - q(D^2+r_2\lambda D) 
+ \frac{1}{2}\left(\frac{\omega_{\rm 2\ rot}}{\omega_{\rm orb}}\right)^2(1+q){r_2^2}(1-\nu^2) \label{eqn:omega2} \\
\end{equation}
where all lengths are in units of the semi-major axis $a$.

\subsection{Solving for the Stellar Surfaces}
The computation of the surface form for each star begins at the pole,
starting with the polar radius
\begin{eqnarray}
r_{\rm 1\ pole} &=&  \sqrt{\frac{G m_1}{g_{\rm 1\ pole}}} \\
r_{\rm 2\ pole} &=&  \sqrt{\frac{G m_2}{g_{\rm 2\ pole}}}
\end{eqnarray}
for the primary and secondary respectively, where $g_{\rm 1\ pole}$ and $g_{\rm 2\ pole}$ are the polar surface gravities
of the components. The primary's mass is obtained from Kepler's third law and the secondary's mass  is obtained from the mass ratio,
\begin{eqnarray}
m_1 &=& \frac{4\pi^2a^3}{GP^2(1+q)} \\
m_2 &=& m_1q
\end{eqnarray}
The values of the simplified potentials at the pole ($\vartheta$ = 0) of
each star from Equations \ref{eqn:omega1} and \ref{eqn:omega2} are
\begin{eqnarray}
\Omega_{\rm 1\ surface} &=& \frac{1}{r_{\rm 1\ pole}} + \frac{q}{\sqrt{D^2 + r_{\rm 1\ pole}^2}} \label{eqn:omega1_surface} \\
\Omega_{\rm 2\ surface} &=& \frac{1}{\sqrt{D^2 + r_{\rm 2\ pole}^2}} + \frac{q}{r_{\rm 2\ pole}} + \frac{1}{2}(1+q)D^2 - qD^2 \label{eqn:omega2_surface} 
\end{eqnarray}

A Newton-Raphson iteration is used to find the radius at the next co-latitude starting from the pole:
\begin{equation}
r_{\rm new} = r_{\rm guess} - \frac{\left(\Omega(r_{\rm guess},\varphi,\vartheta) - \Omega_{\rm surface}\right)}{\partial\Omega/\partial r}
\end{equation}
where $\Omega(r_{\rm guess},\phi,\theta)$ is the value computed from
Equation \ref{eqn:omega1} in the case of the primary or
Equation \ref{eqn:omega2} in the case of the secondary and $\Omega_{\rm
surface}$ corresponds to the constant (for each epoch) from
Equation \ref{eqn:omega1_surface} or Equation \ref{eqn:omega2_surface}.  
The value of 
$r_{\rm new}$ becomes $r_{\rm guess}$ for the next iteration until the
difference $r_{\rm new} - r_{\rm guess}$ is less than $10^{-6}$.  In
practice the iteration always converges to a difference just less than
$10^{-4}$.

The simplified potential derivatives with respect to the radius vector are
\begin{eqnarray}
\frac{\partial\Omega_1}{\partial r_1} &=& -\frac{1}{r_1^2} - \frac{q(r_1-\lambda D)}{(D^2 + r_1^2 - 2r_1\lambda D)^{3/2}} 
- q\lambda D + \left(\frac{\omega_{\rm 1\ rot}}{\omega_{\rm orb}}\right)^2 (1+q)r_1(1-\nu^2) \\
\frac{\partial\Omega_2}{\partial r_2} &=& -\frac{D\lambda + r_2}{(D^2 + r_2^2 + 2r_2\lambda D)^{3/2}} - \frac{q}{r_2^2} +
\lambda D +  \left(\frac{\omega_{\rm 2\ rot}}{\omega_{\rm orb}}\right)^2 (1+q)r_2(1-\nu^2).
\end{eqnarray}
The surface gravity at each point on the surface is most easily computed
from components in rectangular coordinates.  The x-component 
of the local effective surface gravity for the primary is
\begin{equation}
g_x = \frac{\partial\Omega_1}{\partial x} = \frac{-x}{(x^2 + y^2 + z^2)^{3/2}} + \frac{q(D-x)}{((D-x)^2 + y^2 + z^2)^{3/2}} + 
\left(\frac{\omega_{\rm 1\ rot}}{\omega_{\rm orb}}\right)^2 (1+q)x - qD\\
\end{equation}
and for secondary,
\begin{equation}
g_x = \frac{\partial\Omega_2}{\partial x} = \frac{-x}{(x^2 + y^2 + z^2)^{3/2}} + \frac{q(D-x)}{((D-x)^2 + y^2 + z^2)^{3/2}} + 
\left[1- \left(\frac{\omega_{\rm 2\ rot}}{\omega_{\rm orb}}\right)^2\right] (1+q)(D-x) + x(1+q) -qD. \\
\end{equation}
The y and z components have the same form for both stars,
\begin{equation}
g_y = \frac{\partial\Omega_{\rm 1,2}}{\partial y} = \frac{-y}{(x^2 + y^2 + z^2)^{3/2}} + \frac{-yq}{((D-x)^2 + y^2 + z^2)^{3/2}} + 
\left(\frac{\omega_{\rm 1,2\ rot}}{\omega_{\rm orb}}\right)^2 (1+q){y}\\
\end{equation}
\begin{equation}
g_z = \frac{\partial\Omega}{\partial z} = \frac{-z}{(x^2 + y^2 + z^2)^{3/2}} + \frac{-zq}{((D-x)^2 + y^2 + z^2)^{3/2}},
\end{equation}
where the effective surface gravity then is
\begin{equation}
g_{\rm eff} = (g_x^2 + g_y^2 + g_z^2)^{1/2}.
\end{equation}
The effective temperature at each point on the surface 
of either star is  then
\begin{equation}
\label{eqn:vonzeipel}
T_{\rm eff}(r_{1,2},\varphi, \vartheta) = T_{\rm eff\ 1,2\ pole} \left(\frac{g_{\rm eff}(r_{1,2},\varphi,\vartheta)}{g_{\rm 1,2\ pole}}\right)^{\beta}.
\end{equation}
where the exponent $\beta = 0.25$ would correspond
to Von Zeipel's purely radiative gravity darkening law.

Following \citet[Equation 2-6, page 175]{Kopal_binary_book}, the
direction cosines of the normal to surface at each point, which differ
from the direction cosines of the radius vector for non-spherical
stars, are
\begin{eqnarray}
  l &=& g_x/g_{\rm eff} \\
  m &=& g_y/g_{\rm eff} \\
  n &=& g_z/g_{\rm eff} 
\end{eqnarray}
From \cite[Equation 2-3, page 174]{Kopal_binary_book}, the
cosine of the angle between the radius vector and the surface normal is
\begin{equation}
\cos \beta = \lambda l + \mu m + \nu n
\end{equation}
which is used below to compute the flux and the cosine of the
emergent radiation relative to the surface normal for the observer at epoch $t$ is 
\begin{equation}
\mu(\varphi,\vartheta,t) = l l_0 + m m_0 +  n n_0
\label{eqn:mu_value}
\end{equation}
where $l_0$, $m_0$, $n_0$ denote the direction cosines for the line of sight:
\begin{eqnarray}
  l_0 &=& -\sin \psi \sin i \\
  m_0 &=& \cos \psi \sin i \\
  n_0 &=& \cos i.
\end{eqnarray}
These direction cosine formulae differ from \cite[Equation 2-5, page 174]{Kopal_binary_book} where 
$\psi$ is measured from the moment of superior conjunction, when
the primary eclipses the secondary for high inclination systems.
We define angle $\psi$ as 
\begin{equation}
\psi  = \omega + \upsilon,  
\end{equation}
the position of the secondary star from the line of nodes,
which is less than \cite{Kopal_binary_book} by 90 degrees.
The angle $\omega$, the longitude of periastron, for each time step $t$ is
\begin{equation}
\omega =  \omega_0+ \left(\frac{360}{365.25}\right)\frac{t-T_0}{U}
\end{equation}
where $\omega_0$ is the longitude of periastron at a reference epoch $T_0$ and
$U$ is the apsidal period.

The true anomaly is calculated following \cite{SSI1971},
\begin{equation}
\tan\frac{\upsilon}{2} = \left(\frac{1+e}{1-e}\right)^{1/2}\tan \frac{E}{2}
\end{equation}
where the eccentric anomaly E from the first order approximation is
\begin{equation}
E = M + \frac{e\sin M}{1-e\cos M}
\end{equation}
and $M$ is the mean anomaly,
\begin{equation}
M=2\pi\frac{t-T_0}{P}.
\end{equation}

\subsection{Interpolation of the Rest Frame Intensities}
Following \cite{Vega06}, the intensity at each point $(\varphi,\vartheta)$ on each stellar surface is interpolated from synthetic radiation fields calculated using the PHOENIX model atmosphere code \citep{H99, H10, H14}. 
For this work we used PHOENIX version 18.02.00A
and the list of atomic line data designated \verb|atomic_lines.20191028|
for LTE lines and model atoms for hydrogen and helium in non-LTE. 
We assume solar abundances for Spica A and B, although
\cite{MOST_2016} found both stars slightly deficient, by $\simeq -0.2$ dex,
in carbon and aluminum relative to the Sun.

These atmosphere models have 64 depth layers in spherical mode with 64 shells, 63 tangent rays 
and 15 core-intersecting rays. Model boundary conditions include the outer pressure of $10^{-5}$ Pa and a continuum 
optical depth of $10^{-10}$ at 5000\,\AA, with an inner continuum optical depth of $10^{2}$.

The model grid for the primary star spans 20,000 K to 26,000 K in
\teff\ and 3.30 to 3.85 in $\log_{10}(g)$,
\begin{eqnarray}
T_{1j}           = 20,000 + 250\cdot j \quad {\rm K} &\quad &j=\{0,1,\ldots,25\} \\
\log_{10}(g_{1l})     = 3.15 + 0.10\cdot l          &\quad &l=\{0,1,\ldots,13\}.
\end{eqnarray}
for a total of 325 primary star models and the grid for
the secondary  star spans 21,000 K to 24,000 K in \teff\ and
3.85 to 4.20 in $\log_{10}(g)$,
\begin{eqnarray}
T_{2j}           = 21,000 + 250\cdot j \quad {\rm K}&\quad &j=\{0,1,\ldots,13\} \\
\log_{10}(g_{2l})     = 3.85 + 0.05\cdot l              &\quad &l=\{0,1,\ldots,8\}.
\end{eqnarray}
for a total of 104 secondary stars models.

At each rest wavelength the intensity, $I(\lambda, \mu)$, is evaluated at 64 angles
for each point on a star corresponding to a local  $T_{\rm
eff}$, $\log_{10}(g_{\rm eff})$, and emergent angle $\mu$ along the line of sight is obtained using the
trivariate interpolation IDL routine {\tt INTERPOLATE}.

\subsection{Mapping Intensities on to the Plane of the Sky}
An intensity $I_\lambda(\varphi,\vartheta)$ originates from a location
$(x,y,z)$ computed from Equations
\ref{eqn:lambda_cosine_radius_vector}-\ref{eqn:z_coord}.  Following
\citet[see Equations 3-13 to 3-16, pages 44-45]{Kopal_binary_book} the
coordinates of this location in the plane of the sky are
\begin{eqnarray}
x_{\rm sky} &=& x\left[-\cos (\omega + \upsilon) \cos \Omega + \cos i\sin\Omega\sin(\omega + \upsilon) \right] \nonumber  \\
            & &+y\left[-\sin (\omega + \upsilon) \cos \Omega - \cos i\sin\Omega\cos(\omega + \upsilon)\right]+z\left[\sin i\sin\Omega\right]\\ 
y_{\rm sky} &=& x\left[\cos (\omega + \upsilon) \sin \Omega + \cos i\cos\Omega\sin(\omega + \upsilon) \right] \nonumber \\
            & &+y\left[\sin (\omega + \upsilon) \sin \Omega + \cos i\cos\Omega\cos(\omega + \upsilon)\right] + z\left[\sin i\cos\Omega\right] 
\end{eqnarray}
where the un-subscripted $\Omega$ is the orbital element, the position
angle of the ascending node, not to be confused with the potentials
above.  This coordinate transformation converts the x, y, and z
locations on each star in the rotating frame, with the x-axis along a
line towards the secondary and the z-axis perpendicular to the orbital
plane, to the plane of the sky where the x-axis points north and the
z-axis points to the observer.  One additional rotation, 90\degr\, counter clockwise, brings the positive x-axis to point east.

After the transformation and rotation, $x_{\rm sky}$ and $y_{\rm sky}$
correspond to offsets in right ascension and declination
($\Delta\alpha$, $\Delta\delta$) relative to the the center of mass of
the primary star in units of milliarcseconds. This mapping produces a
synthetic image of each star at each epoch $t$. Next, the
intensity map of each star is re-gridded (using the IDL routine {\tt
TRIGRID}) from the original sampling based on the $\varphi,\vartheta$
grid onto a regular 512 $\times$ 512 grid of points in $x_{\rm sky}$
and $y_{\rm sky}$.  Next, the two images are added to form a synthetic
image of the binary.

\subsection{Computation of the Observer's Frame Spectrum}
The projected velocities of a surface element on each the two stellar components are
\begin{equation}
v_1(r_1,\varphi,\vartheta) = -a r_1(\varphi,\vartheta)(\omega_{\rm 1\ rot}) (\lambda m_0 - \mu l_0) + K_1 \left(e \cos \omega + \cos (\upsilon + 
\omega)\right)\\
\end{equation}
and
\begin{equation}
v_2(r_2,\varphi,\vartheta) = -a r_2(\varphi,\vartheta) (\omega_{\rm 2\ rot}) (\lambda m_0 - \mu l_0) + K_2 \left(e \cos (\omega + \pi) + \cos (\upsilon + \omega + \pi)\right)
\end{equation}
where the radii $r_1$ and $r_2$ are in units of the semi-major axis $a$.  The first term 
follows \citet[see Equation 1]{wilson_sofia76}  except we do not  
assume synchronization, so $\omega_{\rm 1,2\ rot} \not= \omega_{\rm orb}$.
The second term, due to the orbital motion of each component, is from
\citet[Equation 197, page 177]{binnendijk_book} where
the projected semi-amplitude velocities for the primary and secondary are
%
\begin{eqnarray}
K_1 &=& \frac{2\pi}{P} \left(\frac{a}{1+q}\right) \frac{\sin i}{(1-e^2)^{1/2}}\\
K_2 &=& \frac{2\pi}{P} \left(\frac{a}{1+(1/q)}\right) \frac{\sin i}{(1-e^2)^{1/2}}
\end{eqnarray}
from \citet[Equation 162, page 151]{binnendijk_book}.  It is assumed
that the rotational angular momentum vectors and orbital angular
momentum vector are parallel such that the stars rotate in the same
sense that they orbit the center of mass.

For each surface element, the observed wavelength is calculated
\begin{equation}
\lambda_{\rm 1,2\ obs} =  (1 + v_{1,2}/c)\lambda_{\rm rest}
\end{equation}
where $c$ is the speed of light in vacuum and $\lambda_{\rm rest}$ is
the vacuum rest-frame wavelength from model atmosphere grid.  The
intensity at a single observed wavelength is the sum of intensities
from a range of rest-frame wavelengths which are Doppler shifted by
rotational and orbital motion.  An observer-frame wavelength grid is
constructed by truncating the rest-frame wavelength grid at both the
blue and red ends such that at every observer-frame wavelength the
contribution of intensities from available rest-frame wavelengths is
complete.  


Next, the intensity at each surface element is interpolated onto the
common observer-frame wavelength grid for both stars. Once this is done, the
sum of the surface integrals \cite[see Equation 35 therein]{linnell84} 
over both stars yield the flux at the Earth at each
observer-frame wavelength
\begin{equation}
F_{1,2}(\lambda_{\rm obs}) = \frac{a^2}{d^2}\int_0^\pi\int_0^{2\pi} I_{1,2}(\lambda_{\rm obs},\vartheta,\varphi)r_{1,2}^2(\varphi,\vartheta) \sin\vartheta \frac{\mu_{1,2}(\varphi,\vartheta)}{\cos\beta_{1,2}(\varphi,\vartheta)}\,\mathrm{d}\varphi\,\mathrm{d}\vartheta\\
\label{eqn:flux}
\end{equation}

\begin{equation}
F_{\rm total}(\lambda_{\rm obs}) = F_{1}(\lambda_{\rm obs}) +  F_{2}(\lambda_{\rm obs})
\label{eqn:flux_total}
\end{equation}
where $d$, the distance from Earth, and the radii $r_{1,2}$ are in units
of the semi-major axis $a$ and where the intensities, with units erg
cm$^{-2}$ s$^{-1}$ {\AA}$^{-1}$ sr$^{-1}$, yield fluxes with units of erg
cm$^{-2}$ s$^{-1}$ {\AA}$^{-1}$.  This double integral is computed using the IDL routine {\tt INT\_TABULATED\_2D} (version 1.6) which first
constructs a Delaunay triangulation of points in the
$\varphi\vartheta$-plane.  For computing the flux 
a $\varphi\vartheta$ grid of 97$\times$97 
was found to be sufficient for 1\% flux accuracy. For
0.1\% flux accuracy more than 400 longitude points are required.

\end{document}